\documentclass{aa}  

\usepackage{graphicx}
\usepackage{import}   
\usepackage{xcolor}  
\usepackage{placeins}
\usepackage{txfonts}
\begin{document} 

\title{Simultaneous J-, H-, K- and L-band spectroscopic observations of galactic Be stars
\thanks{Based on observations obtained at the international Gemini Observatory, a program of NSF’s NOIRLab, which is managed by the Association of Universities for Research in Astronomy (AURA) under a cooperative agreement with the National Science Foundation on behalf of the Gemini Observatory partnership: the National Science Foundation (United States), National Research Council (Canada), Agencia Nacional de Investigación y Desarrollo (Chile), Ministerio de Ciencia, Tecnología e Innovación (Argentina), Ministério da Ci\^{e}ncia, Tecnologia, Inova\c{c}\~{o}es e Comunica\c{c}\~{o}es (Brazil), and Korea Astronomy and Space Science Institute (Republic of Korea). This paper includes data gathered with the 6.5 meter Magellan Telescopes located at Las Campanas Observatory, Chile.}
}
\subtitle{I. IR atlas}
\titlerunning{Simultaneous J-, H-, K- and L-band spectroscopic observations of galactic Be stars}

\author{Y.R. Cochetti\inst{1,2}\fnmsep\thanks{Fellow of CONICET}
		\and
		M.L. Arias\inst{1,2}\fnmsep\thanks{Member of the Carrera del Investigador Científico, CONICET}
		\and
		L.S. Cidale\inst{1,2}\fnmsep$^{\star\star\star}$
		\and
		A. Granada\inst{3}\fnmsep$^{\star\star\star}$
		\and
		A.F. Torres\inst{1,2}\fnmsep$^{\star\star\star}$
		}

\institute{Instituto de Astrofísica de La Plata (CCT La Plata - CONICET, UNLP), Paseo del Bosque S/N, La Plata, B1900FWA, Buenos Aires, Argentina\\ \email{cochetti@fcaglp.unlp.edu.ar}
   		\and
        Departamento de Espectroscopía, Facultad de Ciencias Astronómicas y Geofísicas, Universidad Nacional de La Plata
        \and
        Centro Interdisciplinario de Telecomunicaciones, Electrónica, Computación y Ciencia Aplicada (CITECCA), Sede Andina, Universidad Nacional de Río Negro, Anasagasti 1463, San Carlos de Bariloche, Río Negro, Argentina
        }

   \date{Received September 15, 1996; accepted March 16, 1997}

 
  \abstract
   {It is already accepted that Be stars are surrounded by circumstellar envelopes, which are mostly compatible with a disc geometry in Keplerian rotation. Their infrared region is characterised by a moderate flux excess and the presence of hydrogen recombination lines that allow us to obtain information on the physical and dynamical structure of different regions inside the disc. Nevertheless, many of the infrared studies available in the literature show low-resolution spectra, or are restricted to a small object sample or describe just an individual object. Some others analyse reduced spectral ranges or just one infrared band.
   }
    {We aim to obtain a more complete characterisation of the properties of the circumstellar environment of Be stars that helps to constrain the theoretical models of the Be phenomenon.
   }
    {Throughout the last decade, we used the spectroscopic facilities at the Gemini and Las Campanas observatories to obtain quasi-simultaneous spectra in the J, H, K, and L bands of a sample of Be stars with medium resolution. 
   }
    {We present near-infrared, medium-resolution spectra of a sample of galactic Be stars with different spectral subtypes and luminosity classes. We measure different parameters of the hydrogen recombination lines from the Paschen, Brackett, Pfund, and Humphreys series, and use them to diagnose physical conditions in the circumstellar environment. We analysed the equivalent-width (EW) ratio between Br$\alpha$ and Br$\gamma$ lines and different diagrams of flux ratios. We also identify lines from He~{\sc i}, C~{\sc i}, N~{\sc i}, O~{\sc i}, Na~{\sc i}, Mg~{\sc i}, Mg~{\sc ii}, Si~{\sc i}, Fe~{\sc i}, and Fe~{\sc ii}. Analysing the EW measurements of particular He~{\sc i}, Mg~{\sc ii}, Fe~{\sc i}, Fe~{\sc ii} and O~{\sc i} lines, we find that for some lines they correlate with the spectral type of the star. Particularly, the emission of the O~{\sc i}~$\lambda\,1.3168\,\mu$m line decreases towards the later spectral types.
    }
   {We present an atlas of 22 Be stars, that covers a wide infrared (IR) spectral range with quasi-simultaneous observations. From a detailed analysis, we define new complementary criteria to Mennickent's classification of Be stars according to their disc opacity. Some objects in our sample present compact thick envelopes, while in others the envelope is extended and optically thin. The correlation between the full widths at half maximum (FWHM) and the peak separation ($\Delta \mathrm{V}$) versus $V\,\sin\,i$ for the Br10, Br$\delta$, and Hu14 lines reveals that the broadening mechanism is rotational. The Ly$\beta$ fluorescence is a key mechanism to explain the intensity of the emission of Mg~{\sc ii} and O~{\sc i} lines. 
   }


   \maketitle
%
\section{Introduction}

Be stars are rapidly rotating, non-supergiant B-type stars, whose spectra show or have shown the H$\alpha$ line in emission \citep{Jaschek1981,Collins1987}. It is already accepted that these objects present circumstellar envelopes, mostly compatible with a disc geometry in Keplerian rotation \citep{Rivinius2013}. The infrared (IR) part of their spectra is characterised by a moderate flux excess and the presence of numerous hydrogen emission lines of the Paschen, Brackett, Pfund, and Humphreys series. These lines are formed in a region more internal than that of the optical H lines, and, for most of them, the photospheric absorption contribution is negligible \citep{SteeleClark2001,Lenorzer2002Atlas}. Thus, they allow us to obtain information on the physical and dynamical structure of different regions within the disc \citep{Hony2000, Lenorzer2002Diagrama, Mennickent2009, Granada2010, Sabogal2017}. 

While different studies devoted to the near-IR spectra of Be stars show low-resolution data, many of them are restricted to a small sample, and some others analyse reduced spectral ranges. For instance, there are only a few studies done in the J band, and they focus mainly on a particular object or a specific spectral line or element \citep{Mathew2012XPer,Mathew2012,Stefl2009}. Also, individual spectral ranges have already been studied for large samples of Be stars. \citet{SteeleClark2001} presented H-band spectroscopy of Be stars with a spectral resolution of R $\simeq$ 3\,000. They reported Brackett and Fe~{\sc ii} lines in emission and, from the analysis of the strength ratio of the higher Brackett lines to Br$\gamma$, were able to distinguish early- from late-type Be stars. Later, \citet{Chojnowski2015} published high-resolution H-band spectroscopy for a great number of Be stars observed with APOGEE. They found that the Br11 emission line is formed preferentially in a circumstellar disc at an average distance of $\sim2.2\,R_{\star}$, while the higher Brackett lines seem to originate in an innermost region. Several emission lines have been identified for the first time, such as C~{\sc i}~$\lambda\,1.6895\,\mu$m, which is also formed in the inner region of the discs. In a later work, \citet{Chojnowski2017} analysed the variation of the emission strength, peak intensity ratio, and peak separation. Their analysis revealed a variety of temporal variability, including the disappearance and appearance of the line emission on different timescales. In the K band, \citet{Hanson1996}, \citet{Clark2000} and \citet{Granada2010} reported Br$\gamma$, Br$\delta$, and Pfund lines in emission together with lines of He~{\sc i} in emission or absorption, and Mg~{\sc ii}, Fe~{\sc ii} and Na~{\sc i} lines in emission. \citet{Clark2000} related the infrared characteristics to the underlying properties of the stars: objects that present He~{\sc i} features in emission or absorption are B3 or earlier; if the star presents Mg~{\sc ii} in emission but no He~{\sc i}, it is between B2 and B4; objects with Br$\gamma$ emission but no evidence of He~{\sc i} or Mg~{\sc ii} are B5 or later.
\citet{Lenorzer2002Atlas} provided an extensive atlas of early type stars, including a number of Be stars, covering the K and L bands, while \citet{Lenorzer2002Diagrama} analysed the H recombination lines of those stars and constructed a diagram of flux ratios of some selected recombination lines: Hu14/Br$\alpha$ and Hu14/Pf$\gamma$. In this diagram, the location of the objects gives information about the density of the emitting gas. After that, \citet{Mennickent2009} presented a classification scheme for Be stars based on the intensity of the L-band hydrogen-emission lines. The objects in each group fall in different regions of Lenorzer's diagram; thus, this classification scheme is probably connected to the density of the disc. \citet{Granada2010} analysed a sample of eight Be stars and classified them with Mennickent's criterion. They found that for group I objects, the equivalent widths (EW) of Br$\alpha$ and Br$\gamma$ lines are similar, while for stars in group II the EW(Br$\gamma$) is much larger (more than five times) than the EW(Br$\alpha$). Besides this, \citet{Mennickent2009} and \citet{Granada2010} reported  emission lines not only in  Br$\alpha$, Pf$\gamma$, and the Humphreys series but also in He~{\sc i} $\lambda\,4.038\,\mu$m and He~{\sc i} $\lambda\,4.041\,\mu$m. \citet{Sabogal2017} showed a sample of L-band Be-star spectra and correlated the infrared features with the optical H$\alpha$ line behaviour.

Despite the broad studies performed in the near-IR (NIR) bands, an analysis of Be stars observed (quasi) simultaneously in the whole range ($0.8-4.5\,\mu$m) is still lacking. In order to obtain a more complete picture of the properties of their circumstellar environment, we collected a great number of NIR spectra of Be stars that cover a wider spectral range than those reported in previous works. These spectra together with the results we obtained from them, will be presented in a series of future papers. In this first work we present a medium resolution (R$\sim$1\,800 and R$\sim$6\,000) spectral atlas of a sample of 22 galactic Be stars, observed quasi-simultaneously in the J, H, K, and L bands. The paper is organised as follows. In Sect.~\ref{observations} we present the observations. In Sect.~\ref{results} we present and describe the spectra and highlight their main features. The discussion is presented in Sect.~\ref{discussion}, and our conclusions are in Sect.~\ref{conclusions}. The circumstellar parameters derived from the lines described here will be presented in the second paper of this series.

\section{Observations}\label{observations}

Medium-resolution NIR spectra were obtained for 22 galactic Be stars with spectral types in the O7.5-A0 range and different luminosity classes. The objects were selected either due to their reported variability \citep{Hubert1998,Mennickent2009,Granada2010} or the fact that they have envelope sizes determined by interferometric techniques \citep{Cochetti2019}. This allows us to analyse the evolution of the circumstellar envelopes, with the aim of performing a more complete analysis by adding IR spectroscopic observations.

The observed Be stars are listed in Table~\ref{tabla:info_objetos}, which provides information about the HD name, alternative name, spectral type (ST), luminosity class (LC), projected rotational velocity ($V\,\sin\,i$), effective temperature (T$_{\mathrm{eff}}$), and surface gravity ($\log g$).

\begin{table*}[h!]
\caption{Fundamental parameters for the programme's Be stars.}\label{tabla:info_objetos}      
\centering
\begin{tabular}{l l c c c c c} 
\hline\hline
HD       & Name             & ST$^{(a)}$& LC$^{(a)}$ & $V\,\sin\,i^{(b)}$ \lbrack km/s\rbrack  & T$_{\mathrm{eff}}^{(b)}$ \lbrack K\rbrack  & $\log g^{(b)}$ \lbrack dex\rbrack  \\
\hline
\object{HD 20336}    & BK Cam           & B2.5      & V          & 341 $\pm$ 23           & 18684$\pm$517 & 3.865$\pm$0.072 \\
\hline
\object{HD 23862}    & 28 Tau           & B8        & V          & 290 $\pm$ 15           & 12106$\pm$272 & 3.937$\pm$0.052 \\
\hline
\object{HD 25940}    & 48 Per           & B3        & V          & 220 $\pm$ 13           & 16258$\pm$582 & 3.572$\pm$0.084 \\
\hline
\object{HD 28497}    & 228 Eri          & B2        & V          & 342 $\pm$ 24           & 26724$\pm$427 & 4.200$\pm$0.046 \\
\hline
\object{HD 29441}    & V1150 Tau        & B2.5      & V          & 274 $\pm$ 42$^{(c)}$   & 18000$^{(g)}$ & 3.8$^{(g)}$ \\
\hline
\object{HD 32991}    & 105 Tau          & B2        & V          & 175 $\pm$ 18$^{(d)}$   & 20900$^{(h)}$ & 3.77$^{(h)}$  \\
\hline
\object{HD 35439}    & $\psi_{01}$ Ori  & B1        & V          & 266 $\pm$ 13           & 22134$\pm$665 & 3.920$\pm$0.087 \\
\hline
\object{HD 36576}    & 120 Tau          & B2        & IV-V       & 266 $\pm$ 13           & 22618$\pm$508 & 3.804$\pm$0.062  \\
\hline
\object{HD 37490}    & $\omega$ Ori     & B3        & V          & 155 $\pm$ 5 $^{(e)}$   & 14492$^{(h)}$ & 3.86$^{(h)}$  \\
\hline
\object{HD 41335}    & V696 Mon         & B3-5      & V          & 376 $\pm$ 26           & 20902$\pm$610 & 3.886$\pm$0.081 \\
\hline
\object{HD 68980}    & MX Pup           & B1.5      & IV         & 152 $\pm$ 8           & 25126$\pm$642 & 3.951$\pm$0.077 \\
\hline
\object{HD 142983}   & 48 Lib           & B3  $^{(f)}$& V $^{(f)}$ & 407 $\pm$ 22         & 17642$\pm$554 & 3.845$\pm$0.080 \\
\hline
\object{HD 155806}   & V1075 Sco        & O7.5      & V            & 105 $\pm$ 16$^{(c)}$ & 34000$^{(i)}$ & 4.0$^{(i)}$ \\
\hline
\object{HD 164284}   & 66 Oph           & B2        & V          & 287 $\pm$ 21           & 21609$\pm$523 & 3.943$\pm$0.068  \\
\hline
\object{HD 171623}   & HR 6977          & A0        & V          & 253 $\pm$ 38$^{(c)}$   &  9853$\pm$291 $^{(j)}$   & 3.91$\pm$0.23 $^{(j)}$  \\
\hline
\object{HD 178175}   & V4024 Sgr        & B2        & V          & 111 $\pm$ 7           & 18939$\pm$286 & 3.489$\pm$0.035 \\
\hline
\object{HD 183656}   & V923 Aql         & B7        & III        & 284 $\pm$ 16           & 12626$\pm$524 & 3.295$\pm$0.081 \\
\hline
\object{HD 187811}   & 12 Vul           & B2.5      & V          & 258 $\pm$ 13           & 18086$\pm$583 & 3.810$\pm$0.083 \\
\hline
\object{HD 191610}   & 28 Cyg           & B2.5      & V          & 318 $\pm$ 22           & 18353$\pm$516 & 3.718$\pm$0.072 \\
\hline                  
\object{HD 209409}   & $o$ Aqr          & B7        & IV         & 282 $\pm$ 20           & 12942$\pm$402 & 3.701$\pm$0.067 \\
\hline
\object{HD 212571}   & $\pi$ Aqr        & B1        & III-IV     & 233 $\pm$ 15           & 26061$\pm$736 & 3.915$\pm$0.088 \\
\hline
\object{HD 217050}   & EW Lac           & B4        & III        & 355 $\pm$ 25           & 17893$\pm$509 & 3.571$\pm$0.070 \\
\hline\hline
\end{tabular}
\tablebib{
(a) CDS database, except other source is indicated;
(b) \citet{Fremat2005}, except other source is indicated;
(c) \citet{Glebocki2005};
(d) \citet{Abt2002};
(e) \citet{Chauville2001}; 
(f) \citet{Zorec2005}. 
(g) using $T_{\text{eff}}$ y $\log g$ from other stars with the same ST; 
(h) \citet{Cochetti2019}; 
(i) \citet{Cox2000}; 
(j) \citet{Cochetti2020}.
}
\end{table*}

Observations were performed between 2010 and 2017 using Gemini Near-Infrared Spectrograph \citep[GNIRS;][]{Elias2006a,Elias2006b}, mounted on the Gemini North telescope, under the following programmes: GN-2010B-Q-2, GN-2012B-Q-56, GN-2016B-Q-83, GN-2017A-Q-84, GN-2017A-Q-89, GN-2017B-Q-81, and GN-2017B-Q-86. In Table~\ref{tabla:info_turno}, we describe the instrumental configurations in longslit or cross-dispersed mode that were used. These configurations provide a resolving power of R\,$\sim$\,1800.

\begin{table*}
\caption{Different GNIRS configurations used on each observing run.}\label{tabla:info_turno}      
\centering
\begin{tabular}{l l l l l l} 
\hline\hline
Year        & Mode    & Camera                    & Disperser & Slit  & Bands \\
\hline
2010        & LS      & short blue (0.15''/pix)   & 32 l/mm   & 0.3'' & H-K   \\
\multicolumn{2}{c}{}  & short red (0.15''/pix)    & 32 l/mm   & 0.3'' & L     \\
\hline
2012        & LS      & short blue (0.15''/pix)   & 32 l/mm   & 0.3'' & H-K   \\
\multicolumn{2}{c}{}  & long red (0.05''/pix)     & 101 l/mm  & 0.1'' & L     \\
\hline
2016/2017   & XD      & short blue (0.15''/pix)   & 32 l/mm   & 0.3'' & J-H-K \\
            & LS      & long red (0.05''/pix)     & 10 l/mm   & 0.1'' & L     \\
\hline
\end{tabular}
\end{table*}

In addition, in 2017, we performed observations with the Folded-port InfraRed Echellette (FIRE) spectrometer at Las Campanas Observatory (LCO), Chile. FIRE spectra were obtained with the high-resolution echellete mode covering the 0.8-2.5 $\mu$m range with a resolving power of R\,$\sim6000$. 

For the data obtained in both telescopes, the reduction procedure is similar. Several ABBA sequences were taken for each target. Flats and comparison arc lamps were taken with each source and telluric-standard star pair. We used the {\sc IRAF}\footnote{IRAF is distributed by the National Optical Astronomy Observatory, which is operated by the Association of Universities for Research in Astronomy (AURA) under cooperative agreement with the National Science Foundation.} software package tasks to extract and calibrate the GNIRS spectra and the IDL pipeline provided by Las Campanas Observatory to process FIRE observations. Data reduction steps consisted of subtracting of the AB pairs, flat-fielding, telluric-correction and wavelength and flux calibrations. 

To perform the telluric and instrumental corrections, we used late-B- or early-A-type telluric standard stars, observed near the object in both time and sky position (air mass). The telluric spectrum was used to divide each programme's object spectrum to obtain a telluric-corrected one. Correction for telluric lines is often a difficult task, and, in many cases, we were not able to cancel them out completely. Thus, some residuals of the telluric lines remained in a few spectra, especially at the end of each band and particularly around $2\,\mu$m (K band). However, this effect does not significantly affect the profile of the emission lines, except for the weakest ones.

The wavelength calibration was performed using a calibration lamp, except in the case of the L-band spectra, where we used the telluric lines. Because of the small number of telluric lines available in the standard star spectrum that can be used as wavelength reference, some L-band spectra show a less precise wavelength calibration. 

To flux-calibrate Gemini spectra, we used the telluric spectra subtracted by the hydrogen absorption lines as flux standards. Finally, the spectra were corrected for heliocentric velocities.

In FIRE spectra there are gaps in the regions between orders; for example, around $2.28\,\mu$m. Table~\ref{table:observations} gives the log of the quasi-simultaneous observations, which includes, for each observed Be star, spectrograph, the observing date and observed bands. For the objects for which the time separation between the two observations is longer than ten days, we checked in the BeSS database if the star spectrum showed variations in the optical range between those dates. In all the cases, the H$\alpha$ profile remained very similar.

\begin{table}
\caption{Log of the quasi-simultaneous observations of Be stars in the J, H, K, and L bands.} \label{table:observations}
\centering
\begin{tabular}{l l c p{0.22cm}p{0.22cm}p{0.22cm}p{0.22cm}  } 
\hline\hline
Name              & Spectrograph  & Obs. date  & \multicolumn{4}{c}{Bands} \\
\hline
BK Cam            & GNIRS       & 2016-11-16      & J & H & K & L  \\
\hline
28 Tau            & GNIRS       & 2016-11-16      &   &   &   & L  \\
                  &             & 2016-11-18      & J & H & K &    \\
\hline
48 Per            & GNIRS       & 2013-01-01      &   & H & K & L  \\
\hline
228 Eri           & GNIRS       & 2017-12-10      & J & H & K &    \\
                  &             & 2017-12-14      &   &   &   & L  \\
\hline
V1150 Tau         & GNIRS       & 2012-12-23      &   & H & K & L  \\
\hline
105 Tau           & GNIRS       & 2017-01-04      & J & H & K & L  \\
\hline
$\psi_{01}$ Ori   & GNIRS       & 2017-12-14      &   &   &   & L  \\
                  &             & 2018-01-04      & J & H & K &    \\
\hline
120 Tau           & GNIRS       & 2017-12-29      & J & H & K &    \\
                  &             & 2018-01-04      &   &   &   & L  \\
\hline
$\omega$ Ori      & GNIRS       & 2018-01-04      & J & H & K & L  \\
\hline  
V696 Mon          & GNIRS       & 2017-12-29      & J & H & K &    \\
                  &             & 2018-01-14      &   &   &   & L  \\
\hline
MX Pup            & FIRE        & 2017-06-03      & J & H & K &    \\
\hline
48 Lib            & GNIRS       & 2017-07-09      & J & H & K &    \\
\hline
V1075 Sco         & FIRE        & 2017-06-04      & J & H & K &    \\
\hline
66 Oph            & GNIRS       & 2017-07-03      & J & H & K & L  \\   
\hline
HD\, 171623       & FIRE        & 2017-06-04      & J & H & K &    \\
\hline
V4024 Sgr         & FIRE        & 2017-06-04      & J & H & K &    \\
\hline
V923 Aql          & GNIRS       & 2017-05-29      &   &   &   & L  \\
                  &             & 2017-07-08      & J & H & K &    \\
\hline
12 Vul            & GNIRS       & 2010-09-15      &   & H & K & L  \\
\hline  
28 Cyg            & GNIRS       & 2017-07-09      & J & H & K & L  \\
\hline      
$o$ Aqr           & GNIRS       & 2017-07-10      & J & H & K &    \\
\hline
EW Lac            & GNIRS       & 2017-12-27      &   &   &   & L  \\
                  &             & 2018-01-11      & J & H & K &    \\
\hline  
$\pi$ Aqr         & FIRE        & 2017-06-04      & J & H & K &    \\
\hline
\multicolumn{3}{l}{Observations per band}                          & 19 & 22& 22 & 15 \\
\end{tabular}
\end{table}

\section{Results}\label{results}

\begin{figure*} 
\centering
\includegraphics[angle=0,width=0.95\textwidth]{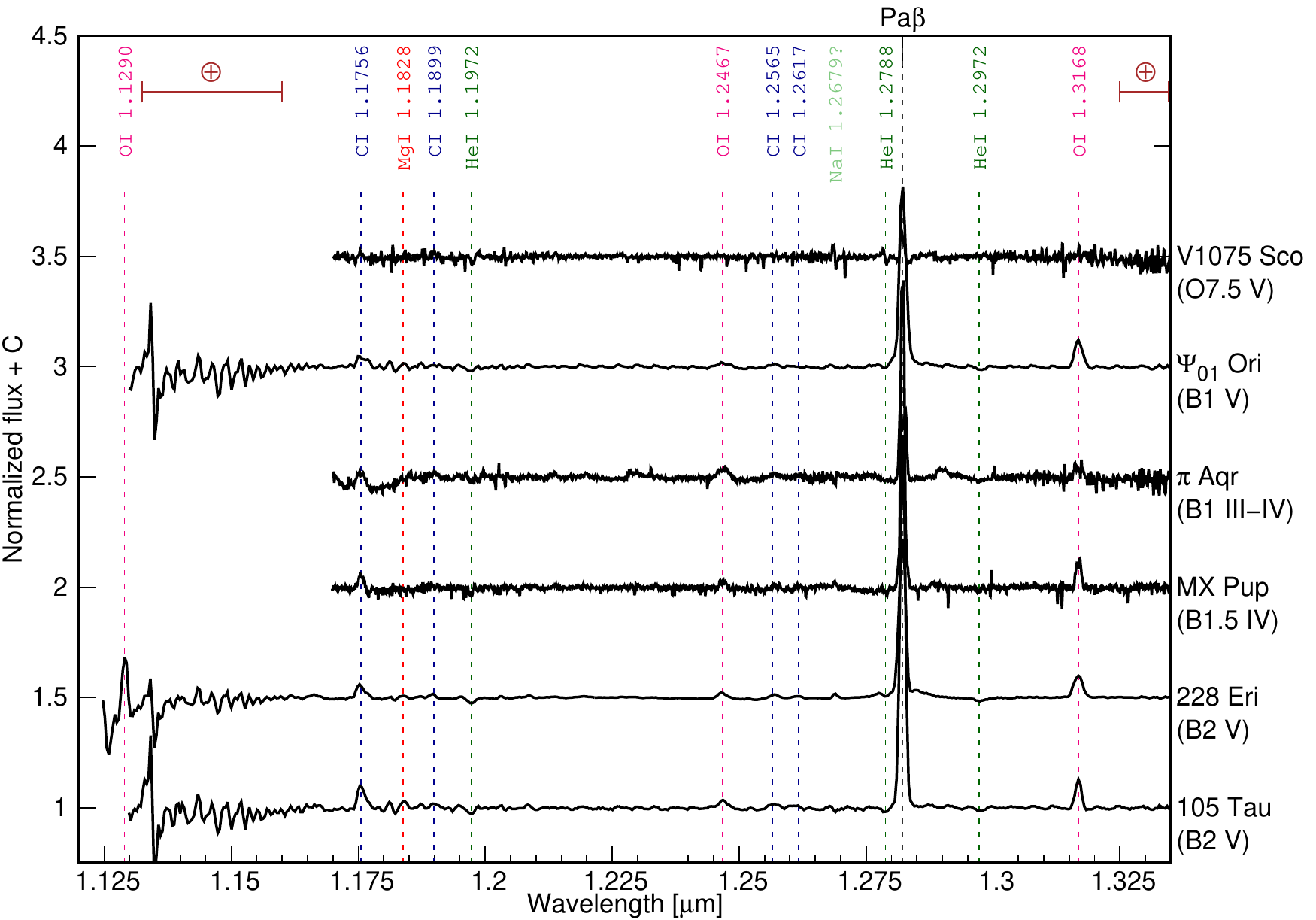}
\includegraphics[angle=0,width=0.95\textwidth]{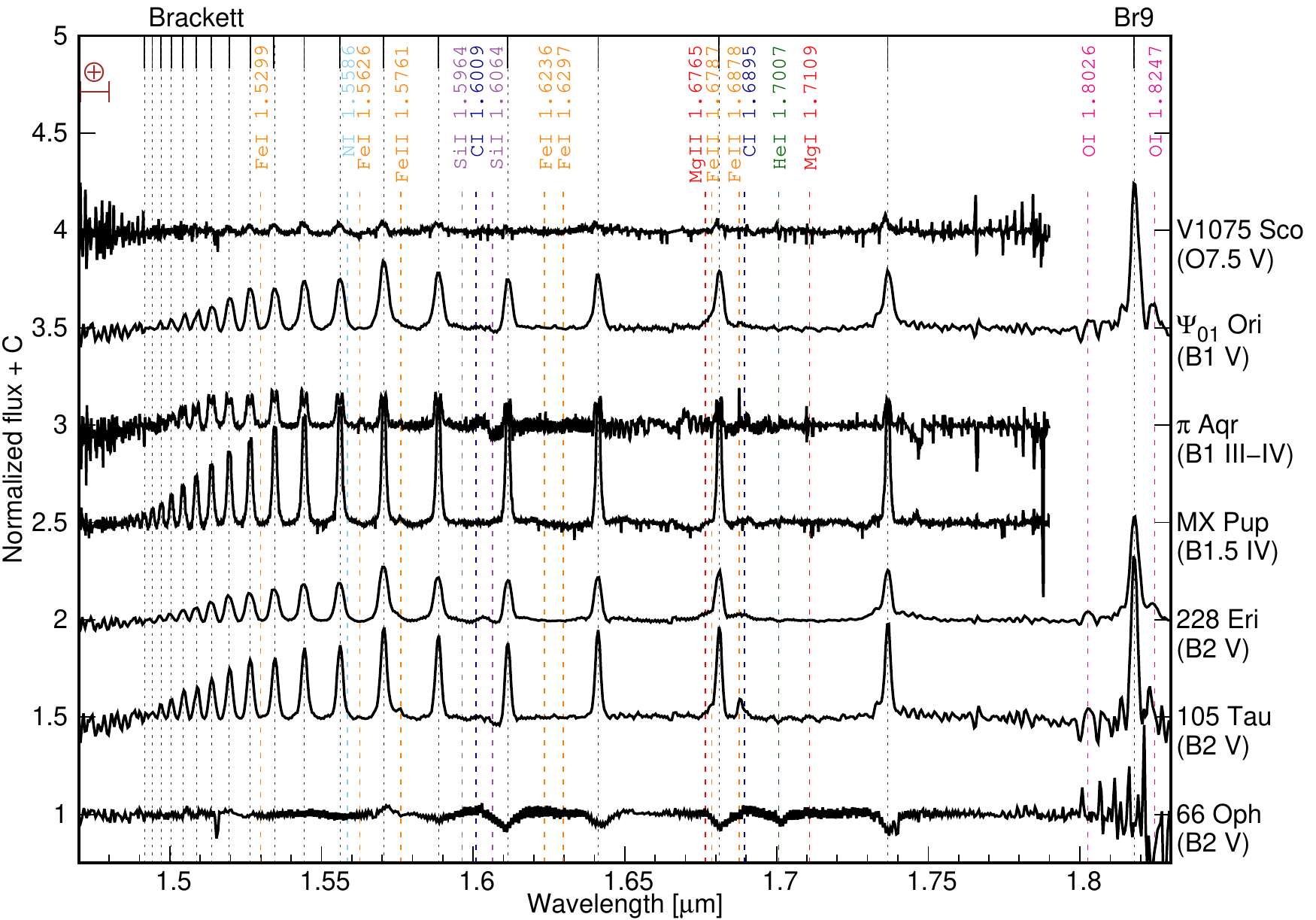}
\caption{J- (upper panel) and H-band (lower panel) spectra of Be stars. Spectra are normalised and shifted vertically to ease comparison.}
\label{fig:AtlasJH}
\end{figure*}

\begin{figure*} 
\centering
\includegraphics[angle=0,width=0.95\textwidth]{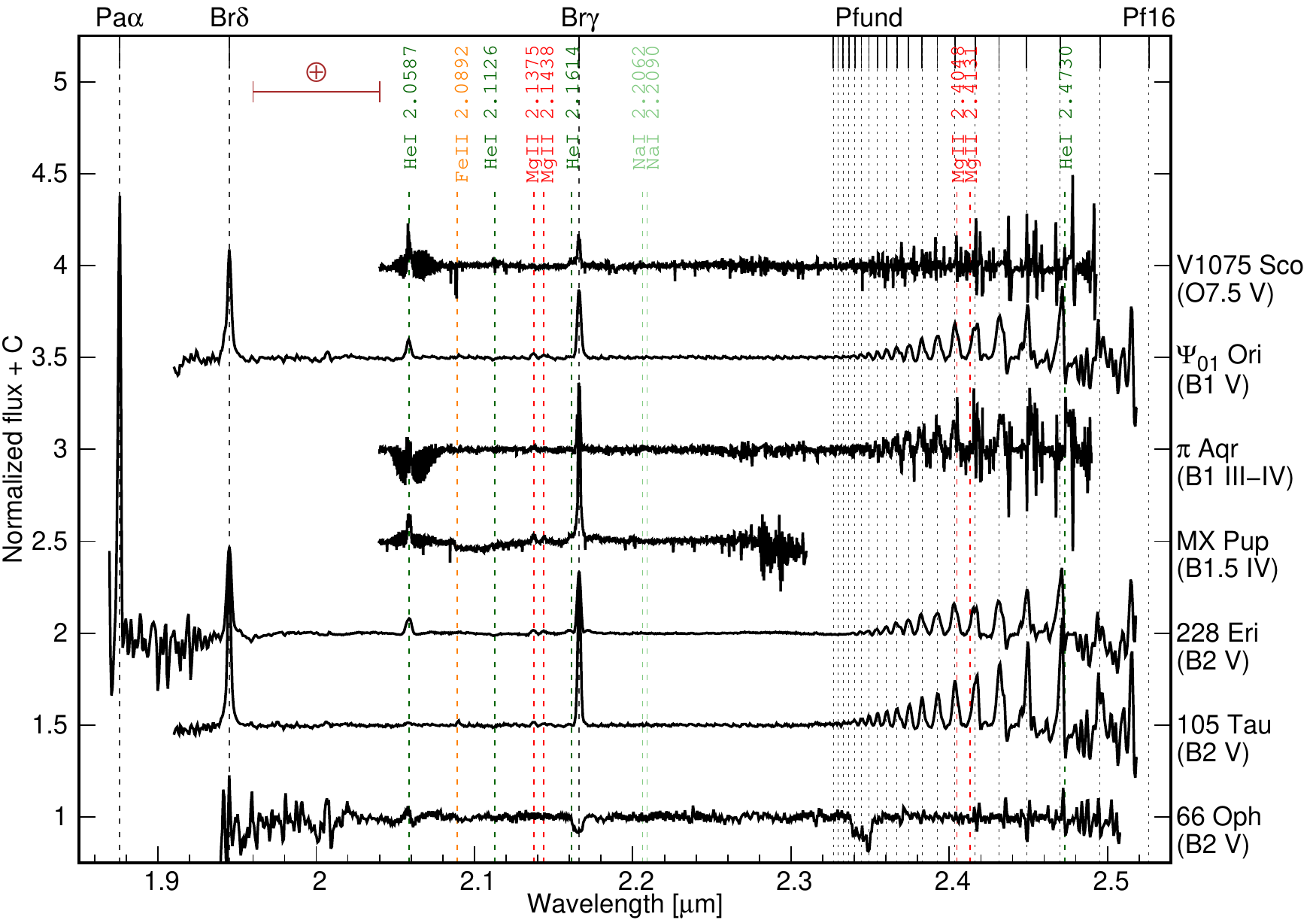}
\includegraphics[angle=0,width=0.95\textwidth]{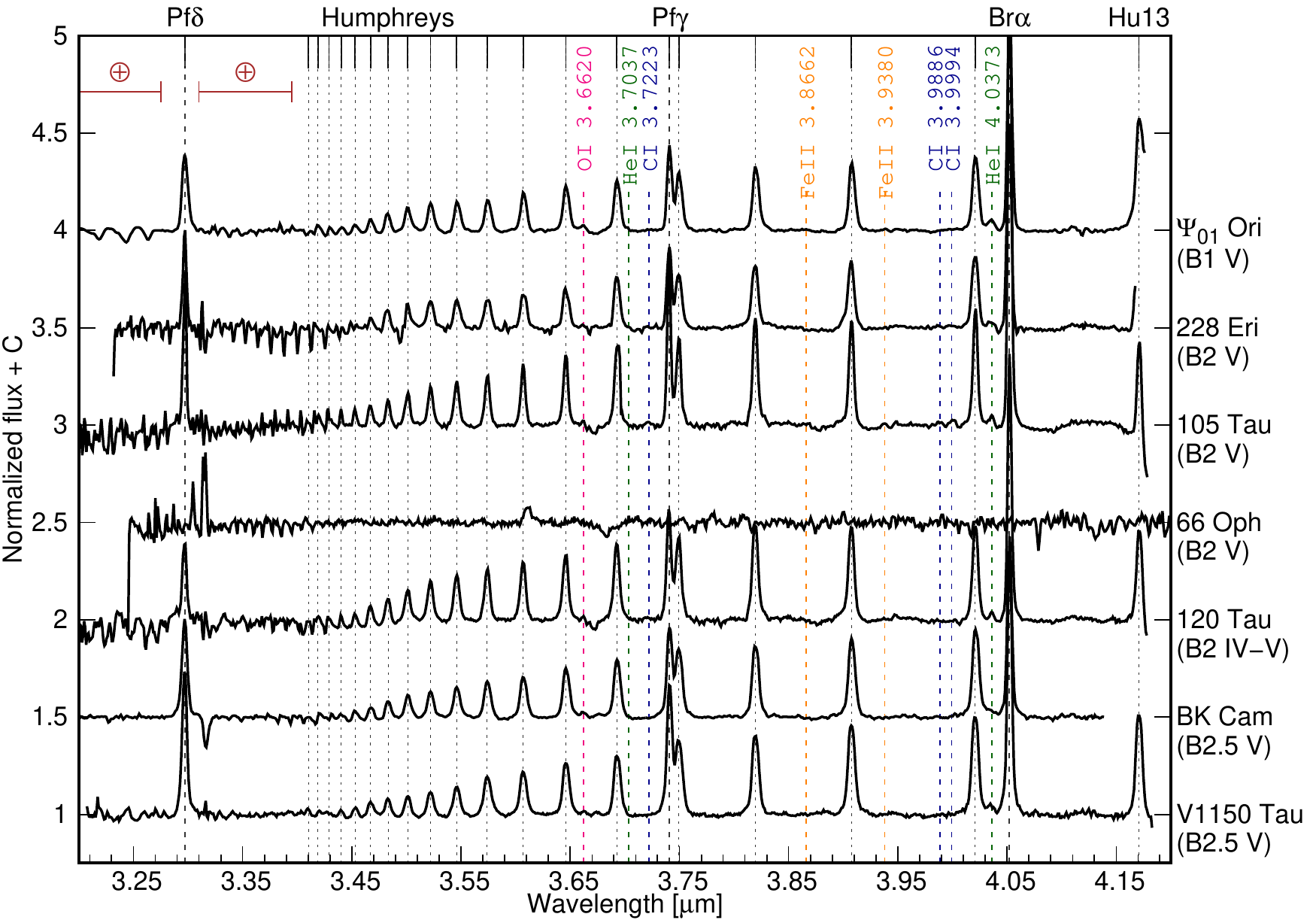}
\caption{K- (upper panel) and L-band (lower panel) spectra of Be stars. Spectra are normalised and shifted vertically to ease comparison.}
\label{fig:AtlasKL}
\end{figure*}

In the following sub-sections, we present and describe the main features that characterise the NIR spectra for the programme's Be stars. Figures~\ref{fig:AtlasJH}, ~\ref{fig:AtlasKL}, and~\ref{fig:AtlasJ2} to ~\ref{fig:AtlasL2} show the complete set of spectra, which were normalised and vertically shifted to ease inspection and comparison. The spectra are organised according to the spectral type of the star, and the positions of the identified features are labelled. The spectral regions with telluric residual lines are marked with the $\oplus$ symbol. Figures~\ref{fig:Paalfa} and \ref{fig:Bralfa} present the line profiles of the most intense lines, Pa$\alpha$ and Br$\alpha$, which are overlapped in the complete spectra figures.

\subsection{Hydrogen recombination lines}\label{Hlines}

Our spectra cover a wide spectral range, where we have identified the following recombination hydrogen (H) lines: Pa$\alpha$ and Pa$\beta$; Br$\alpha$, and Br$\gamma$ up to the end of the series; Pf$\gamma$, Pf$\delta$, and Pf16 up to the end of the series; and Hu13 up to the end of the series. The H lines present different types of profiles. Most of the stars show H lines in emission above the continuum level, with one peak (e.g. $\Psi_{01}$\,Ori) or two peaks (e.g. 12 Vul in the K band). There are some cases where the emission is weak and does not completely fill the photospheric absorption. In these cases the emission could be overimposed on the photospheric absorption (e.g 12\,Vul in the H-band), resembling a shell-type profile (e.g. 28\,Tau in the Brackett series) or barely visible as a deformation in the photospheric absorption-line profile (e.g. HD\,171623 in the Pa$\beta$ and Br$\gamma$ lines). Spectra with no evidence of H emission lines are also observed (e.g. 66\,Oph). 

Tables \ref{tabla:V1075Sco_Hlines} and \ref{tabla:PsiOri_Hlines} to \ref{tabla:28Tau_Hlines} list the obtained line parameters for each object (except for 66\,Oph and HD\,171623, which present no evidence of H emission lines): EW, fluxes (Fl), full widths at half maximum (FWHM) and the peak separation of double-peaked line profiles ($\Delta \mathrm{V}$). The last two values are given in units of velocity. To obtain these parameters we performed Gaussian fittings to the emission components using IRAF tasks. The uncertainties for the measurements are around 10\% for EW and Fl, 75\,km\,s$^{-1}$ for FWHM, and 25\,km\,s$^{-1}$ for $\Delta \mathrm{V}$. In this paper, we employ the convention that positive EW indicate emission features and negative EW indicate absorption features.

EWs and Fl for the H lines are corrected for photospheric absorption. For this purpose, we generated synthetic spectra with the {\sc synspec} program \citep{Hubeny1988,Hubeny1995,Hubeny2011,Hubeny2017} using \citet{Kurucz1979}'s atmosphere models as input. We used models with solar abundances, a microturbulence velocity of 2 km\,s$^{-1}$, and T$_{\mathrm{eff}}$ and $\log g$ in the ranges of $10\,000-39\,000$ K and $2.5-5$ dex, respectively. Steps for T$_{\mathrm{eff}}$ and $\log g$ were $1\,000$ K and $0.5$ dex, respectively. Table~\ref{table:modelos} summarises the selected $\log g$ range for each T$_{\mathrm{eff}}$ interval. Then, we measured the theoretical EWs of the H lines.

\begin{table}
\caption{Selected $\log g$ range for each T$_{\mathrm{eff}}$ interval.} \label{table:modelos}
\centering
\begin{tabular}{c c} 
\hline\hline
T$_{\mathrm{eff}}$ interval& $\log g$ range \\
 \lbrack kK\rbrack & (dex) \\
\hline
10 - 11 & 2.0 - 5.0 \\
12 - 19 & 2.5 - 5.0 \\
20 - 26 & 3.0 - 5.0 \\
27 - 31 & 3.5 - 5.0 \\
32 - 39 & 4.0 - 5.0 \\
\hline
\end{tabular}
\end{table}

\begin{figure} 
\centering
\includegraphics[width=0.45\columnwidth]{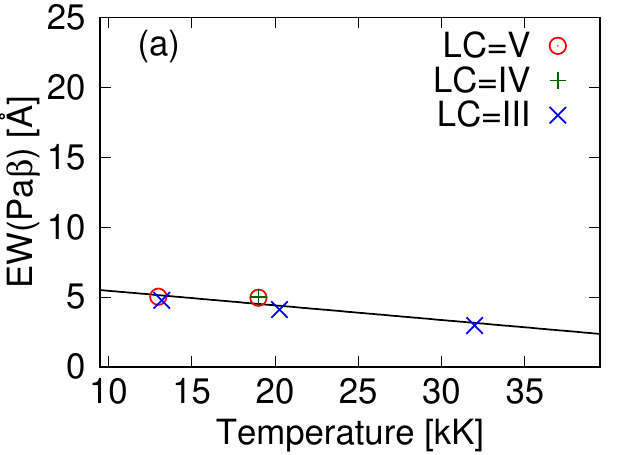}
\includegraphics[width=0.45\columnwidth]{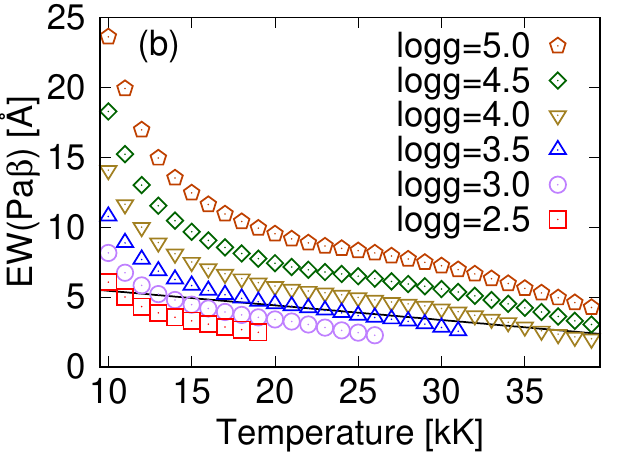}
\includegraphics[width=0.45\columnwidth]{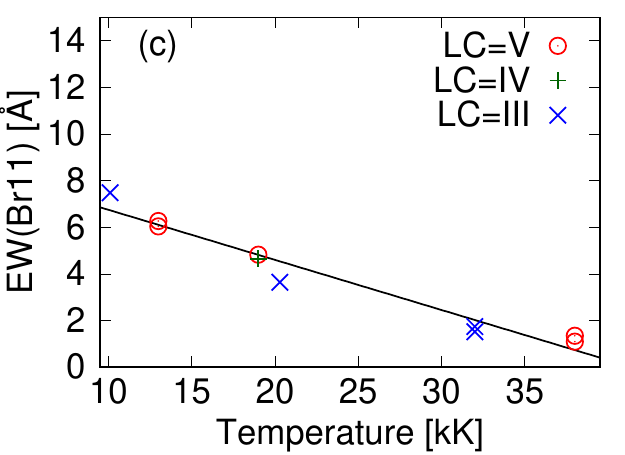}
\includegraphics[width=0.45\columnwidth]{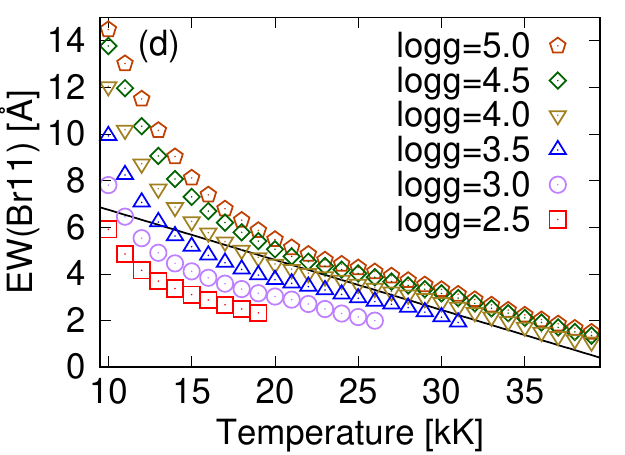}
\includegraphics[width=0.45\columnwidth]{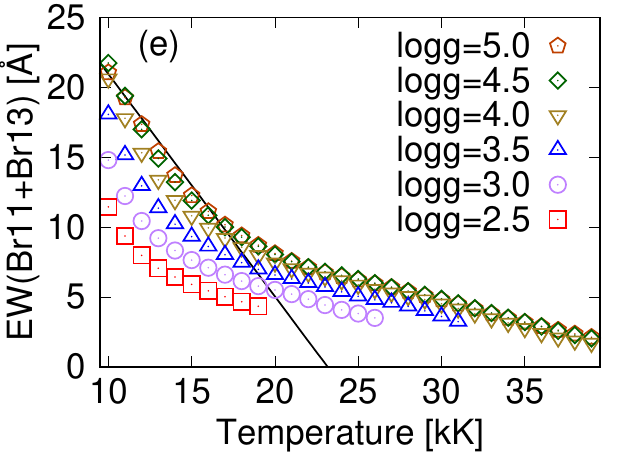}\\
\includegraphics[width=0.45\columnwidth]{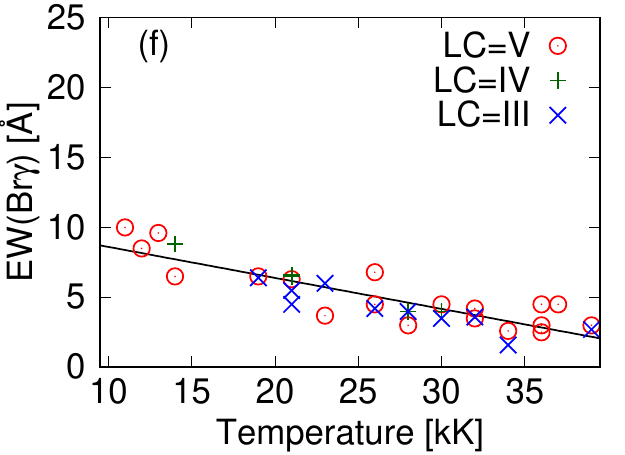}
\includegraphics[width=0.45\columnwidth]{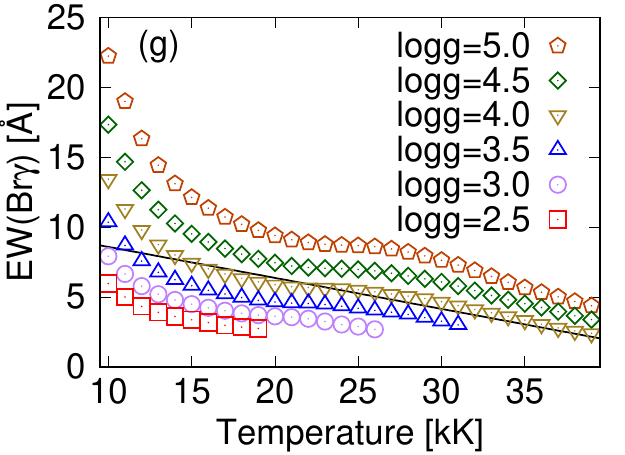}
\includegraphics[width=0.45\columnwidth]{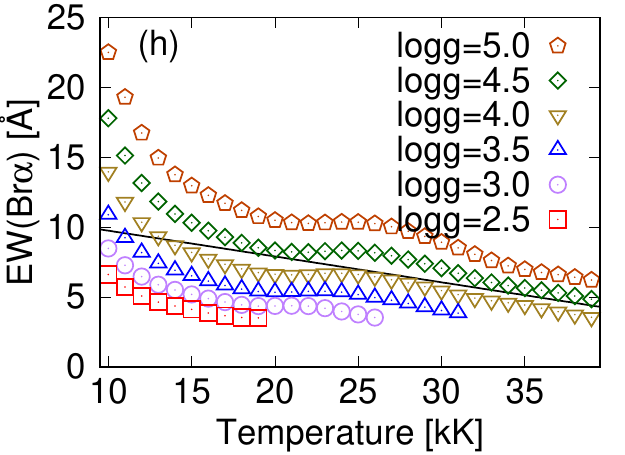}
\includegraphics[width=0.45\columnwidth]{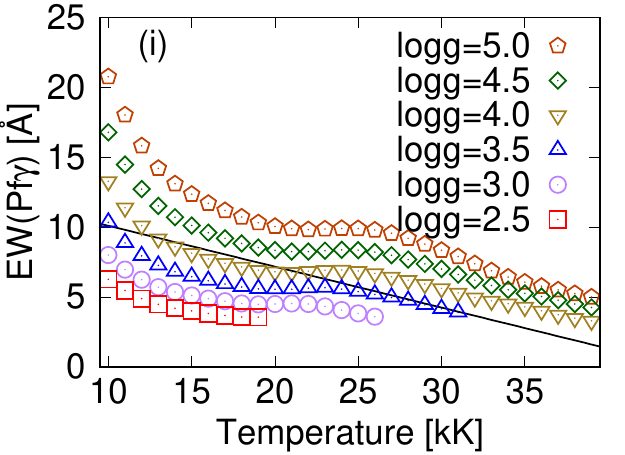}
\includegraphics[width=0.45\columnwidth]{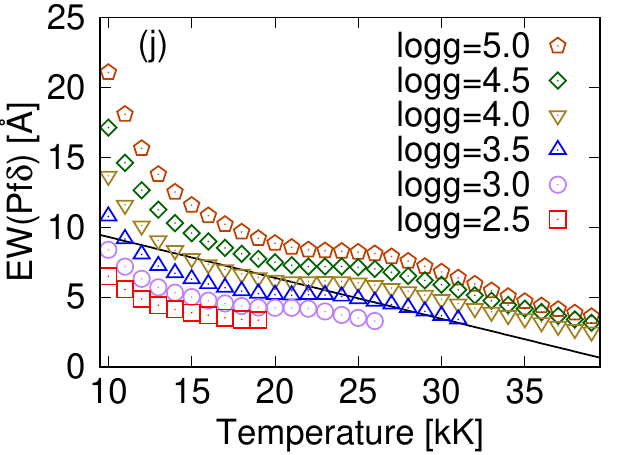}
\caption{Comparison between EW obtained through {\sc synspec}'s spectra and values/fits from the literature (see Sec.~\ref{Hlines} for details).}
\label{fig:corr_photos}
\end{figure}

We checked if the theoretical EWs are in good agreement with those previously reported for B stars. For the Pa$\beta$ lines we made a fitting of the T$_{\mathrm{eff}}$ versus EW behaviour using the values reported by \citet{Wallace2000} (Fig.~\ref{fig:corr_photos} a), and compared it with the EW from the synthetic spectra (Fig.~\ref{fig:corr_photos} b). The fitting was done using an implementation of the non-linear least-squares (NLLS) provided by {\sc GNUPLOT}. In a similar way, Fig.~\ref{fig:corr_photos} c presents the reported values of the EW of Br11 line from \citet{Meyer1998} for OB stars with luminosity classes V-III, and the comparison is in Fig.~\ref{fig:corr_photos} d. Using a bigger sample from the APOGEE-2 DR14 catalogue \citep{Abolfathi2018}, \citet{RamirezPreciado2020} found a linear relationship between the sum of the EWs of the Br11 and Br13 lines and T$_{\mathrm{eff}}$. This relationship, which is shown by a solid line in Fig.~\ref{fig:corr_photos} e, falls in the upper limit of the theoretical EW values for stars with T$_{\mathrm{eff}}$ in the $10\,000-20\,000$~K range.

In the case of the Br$\gamma$ line, we used a fitting obtained from the measurements given by \citet{Hanson1996} (Figs.~\ref{fig:corr_photos} f and g). Figures \ref{fig:corr_photos} h, i and j present the comparison of the EWs from the synthetic spectra with the fittings given by \citet{Lenorzer2002Atlas}. These fittings that give the relationship between the spectral type and the EW of some Brackett and Pfund lines were obtained empirically from a sample of 15 B-type dwarf to giant stars. 

In general, we observe that the EWs obtained from the synthetic spectra are in good agreement with the values previously reported, especially in the T$_{\mathrm{eff}}$ and $\log g$ ranges typically observed in Be stars \citep[$20\,000-30\,000$ K, $3.5-4.5$ dex, e.g. ][]{Cochetti2020}. Then, we used the theoretical EWs to correct the observed H lines in our sample.

\begin{table}[t!]
\caption{V1075\,Sco - EW, Fl, and FWHM of the hydrogen lines.}\label{tabla:V1075Sco_Hlines}  
\centering
\begin{tabular}{l|ccc} 
\hline\hline
Band/line &  EW     & Fl    & FWHM \\  
\hline      
J band\\  
Pa$\beta$ & 4.44  & 535.85 &  197 \\
\hline
H band\\  
Br$_{10}$ & 3.79 & 157.54 & 313\\
Br$_{11}$ & 3.07 & 144.84 & 438\\
Br$_{12}$ & 2.31 & 115.05 & 529\\
Br$_{13}$ & \multicolumn{3}{c}{em}  \\  
Br$_{14}$ & 1.81 & 104.26 & 502\\
Br$_{15}$ & 1.97 & 116.7& 478\\
Br$_{16}$ & 1.7  &103.57 & 509\\
Br$_{17}$ & 1.7 & 104.69 & 540\\
Br$_{18}$ & 1.33 &  83.74& 521\\
Br$_{19}$ & 1.2 & 77.31& 587\\
\hline
K band\\   
Br$\gamma$ & 6.2 & 121.38 & 244\\
\hline
\end{tabular}
\tablefoot{
Positive EW indicate emission features. The EW are in $\AA$, Fl are in unit of 10$^{-13}$\,erg\,cm$^{-2}$\,s$^{-1}\,\AA^{-1}$, and FWHM are in km\,s$^{-1}$.
}
\end{table}

\subsection{Group classification}

\citet{Mennickent2009} proposed a criterion to classify the Be stars into three groups, based on the relative intensity of the hydrogen emission lines observed in the L band. Group I encompasses stars with Br$\alpha$ and Pf$\gamma$ lines with similar intensities to Humphreys lines. Stars with Br$\alpha$ and Pf$\gamma$ lines more intense than Humphreys lines belong to group II, while group III comprises stars with no emission lines. Following this criterion, we classified the stars of our sample with L-band observations (see Figs.~\ref{fig:AtlasKL} and \ref{fig:AtlasL2}) in the groups of Table~\ref{tabla:grupos}.

\begin{table}[t!]
\caption{Group classification of our star sample according to Mennickent's criterion.}\label{tabla:grupos}  
\centering
\begin{tabular}{ccc} 
\hline\hline
Group I   &  Group II        & Group III \\  
\hline      
BK\,Cam   & $\Psi_{01}$\,Ori$^*$ & 66\,Oph \\
12\,Vul   & 228\,Eri$^*$         &         \\
28\,Cyg   & 105\,Tau         &         \\
V696\,Mon & 120\,Tau$^*$         &         \\ 
          & V1150\,Tau$^*$       &         \\
          & 48\,Per          &         \\
          & $\omega$\,Ori    &         \\
          & EW\,Lac          &         \\
          & V923\,Aql        &         \\
          & 28\,Tau          &         \\
\hline
\multicolumn{3}{l}{$^{*}$ Reclassified to group I (see Sec.\ref{criteria}).} \\
\end{tabular}
\end{table}

\subsection{EW(Br$\alpha$)/EW(Br$\gamma$) ratio}

According to the star sample analysed by \citet{Granada2010}, group I objects present Br$\alpha$ and Br$\gamma$ lines with similar EW values up to EW(Br$\alpha$)$\simeq$3\,EW(Br$\gamma$), while for stars in group II, the EW(Br$\alpha$) is much larger (more than five times) than the EW(Br$\gamma$). This behaviour is characteristic of the optically thick models, which predict a strong depression in the emission of Br$\alpha$ \citep{Lynch2000}. 

\begin{figure} 
\centering
\includegraphics[angle=270,width=\columnwidth]{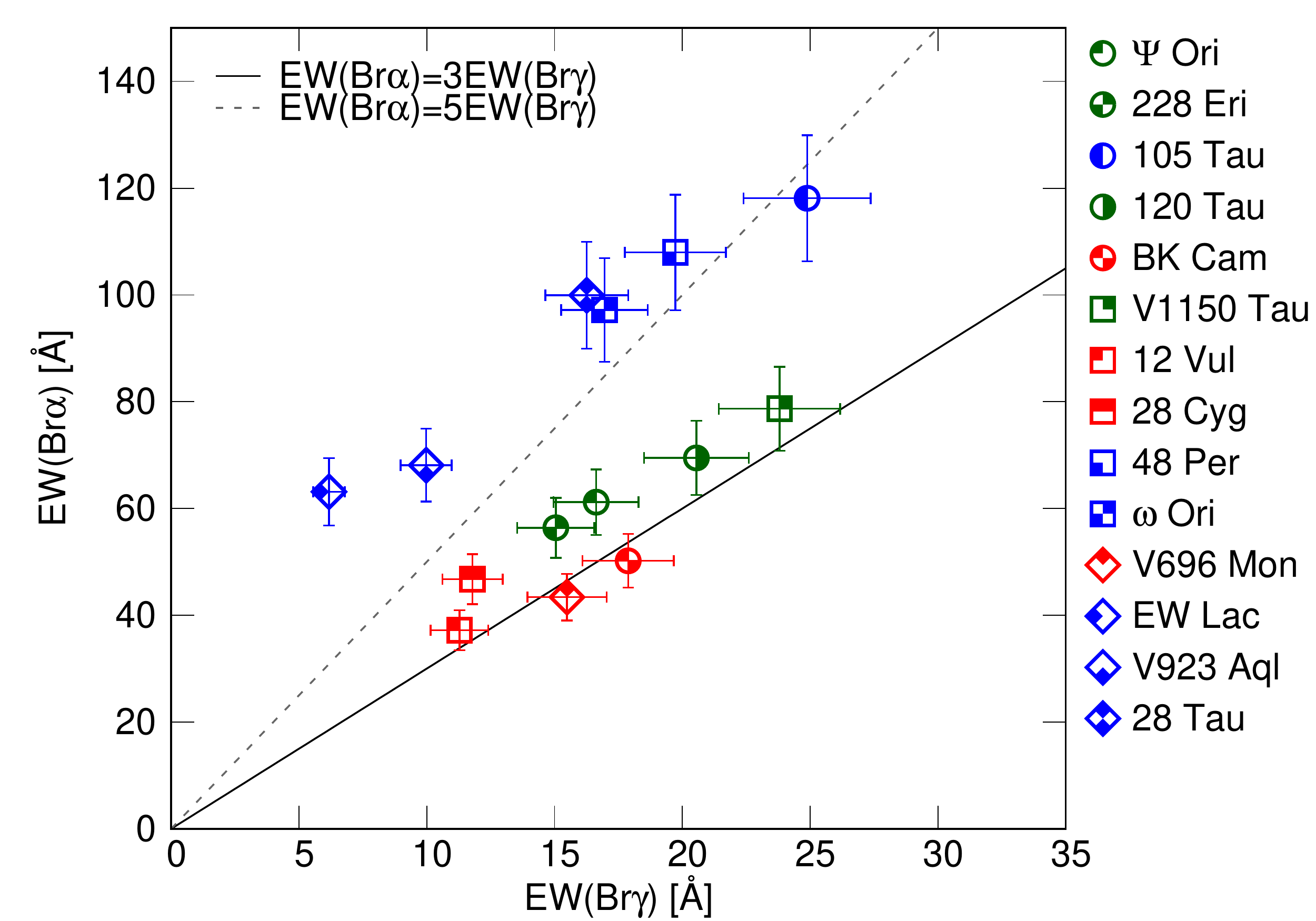}
\caption{EW(Br$\alpha$) versus EW(Br$\gamma$). Red symbols are for group I stars, group II stars are plotted in blue, and green symbols are for the intermediate group I-II. The relations EW(Br$\alpha$)=3\,EW(Br$\gamma$) and EW(Br$\alpha$)=5\,EW(Br$\gamma$) are plotted in continuous and dashed lines, respectively.}
\label{fig:EWBraBrg}
\end{figure}

In Fig.~\ref{fig:EWBraBrg} we plot EW(Br$\alpha$) versus EW(Br$\gamma$) for the programme's stars. Each star is represented with a different symbol. The empirical limits EW(Br$\alpha$)=3\,EW(Br$\gamma$) and EW(Br$\alpha$)=5\,EW(Br$\gamma$) are plotted in continuous and dashed lines, respectively, as a reference. Group I stars, which follow the same trend as the sample of \citet{Granada2010}, are plotted in red. Group II stars are plotted in blue or green symbols according to the following criterion: those which follow the trend from \citet{Granada2010} EW(Br$\alpha$)$\gtrsim$5\,EW(Br$\gamma$) are plotted in blue, and those with the EW(Br$\alpha$)/EW(Br$\gamma$) ratio similar to group I stars are plotted in green. We identify the green-symbol stars as group I-II. It is worth mentioning that 28\,Cyg presents an EW(Br$\alpha$)/EW(Br$\gamma$) ratio a little higher than expected for a group I object.
 
\subsection{Location in the Lenorzer diagram and other flux-ratio diagrams}

\begin{figure*} 
\centering
\includegraphics[angle=270,width=0.9\columnwidth]{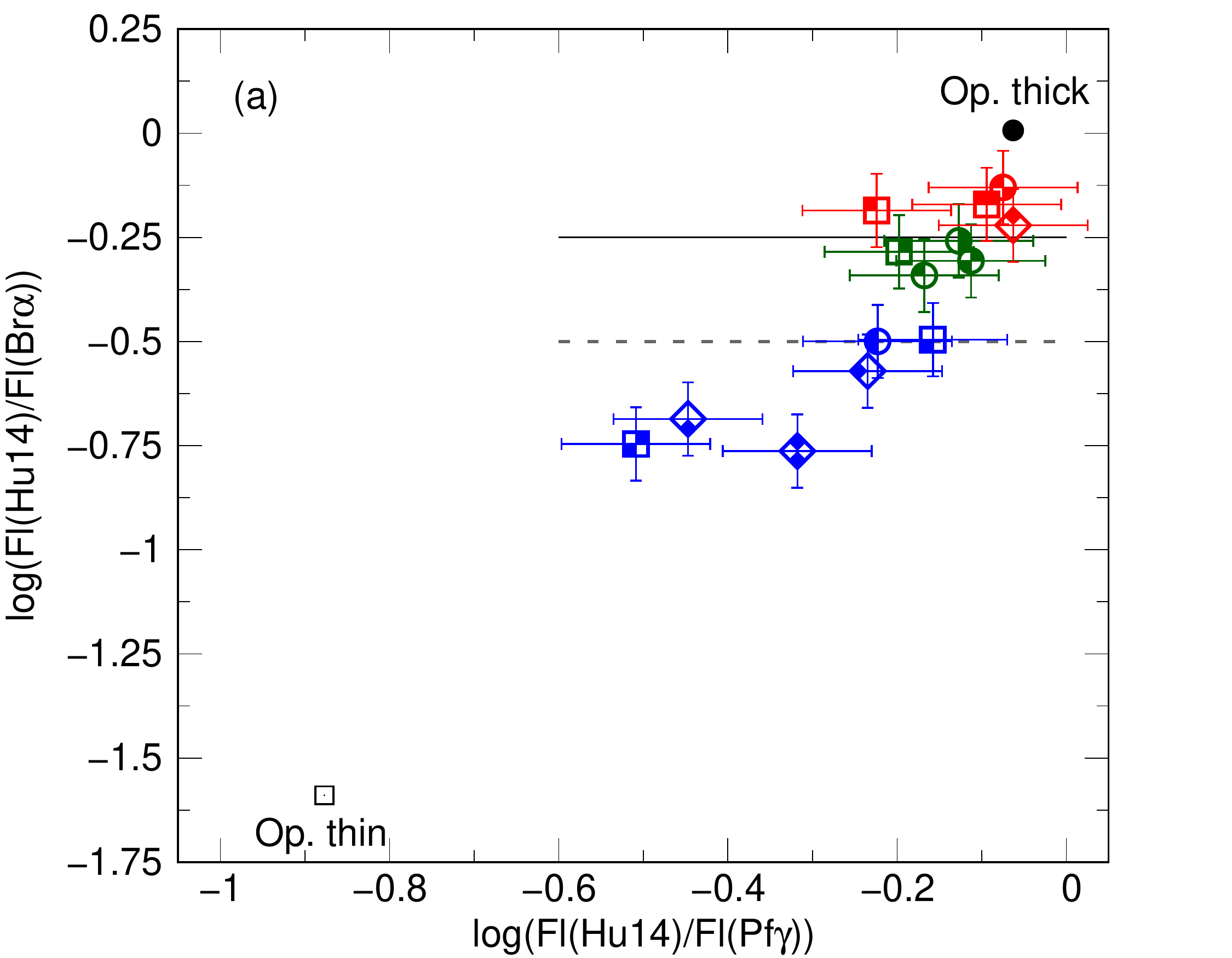}
\includegraphics[angle=270,width=\columnwidth]{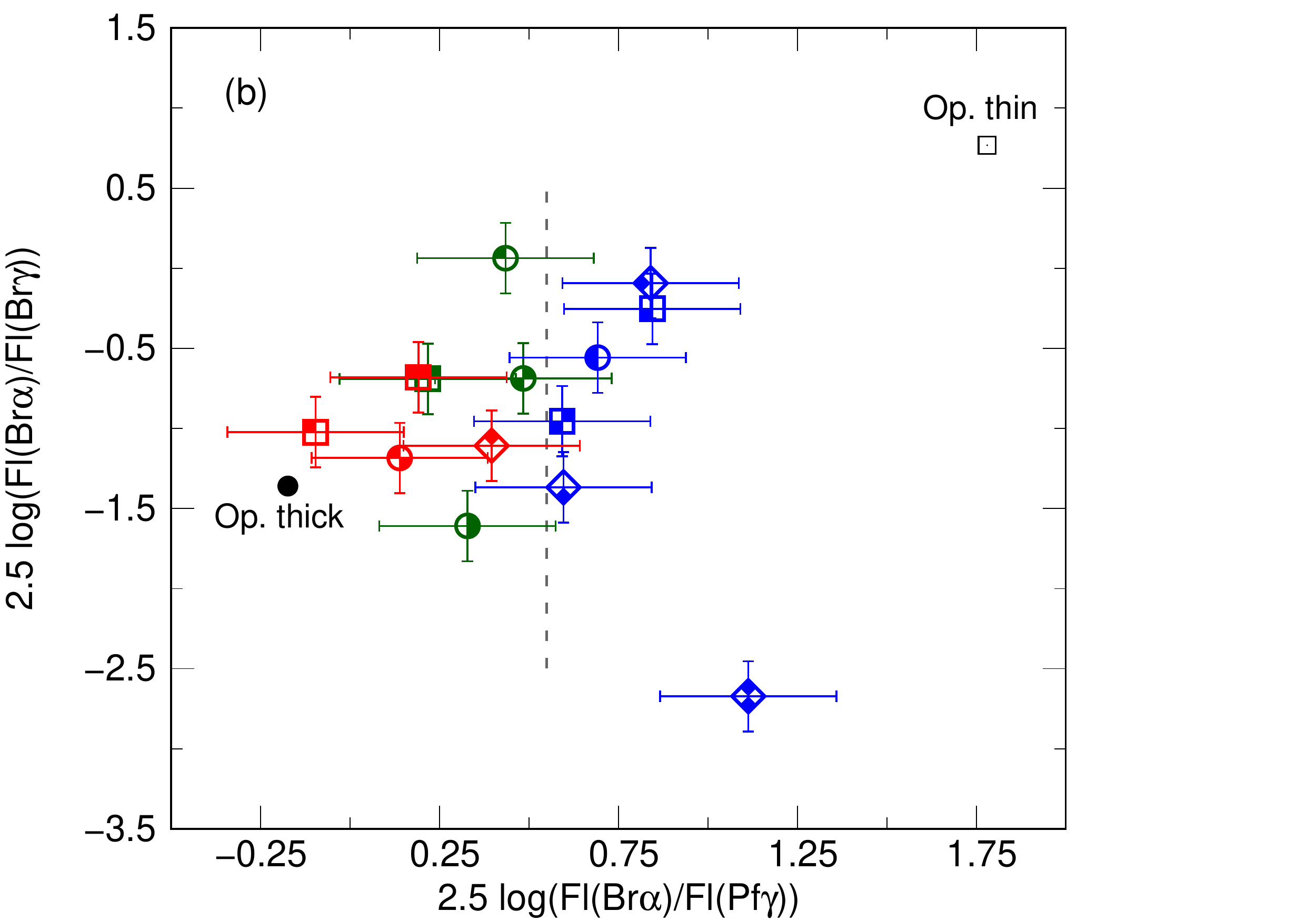}\\
\includegraphics[angle=270,width=0.9\columnwidth]{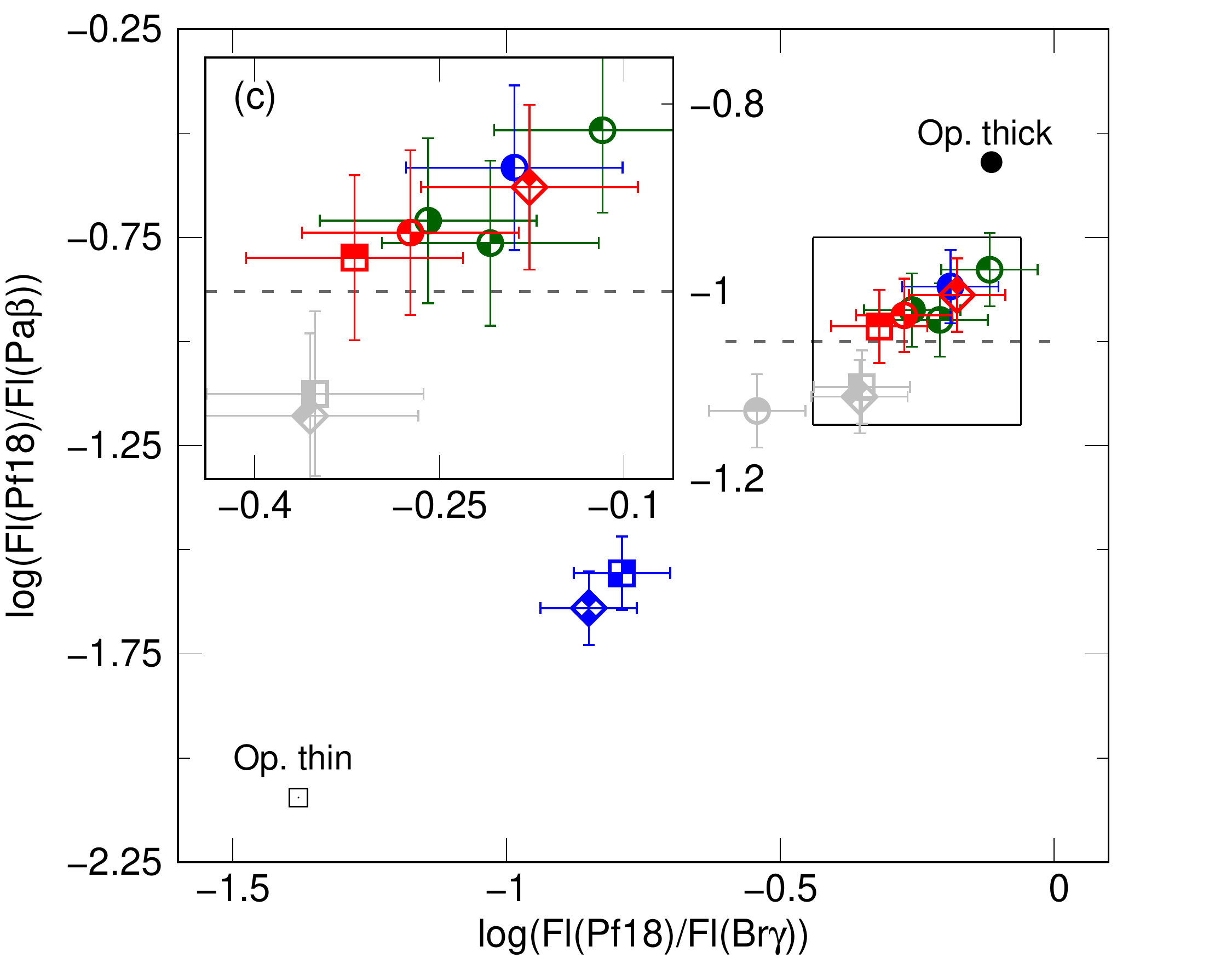}
\includegraphics[angle=270,width=\columnwidth]{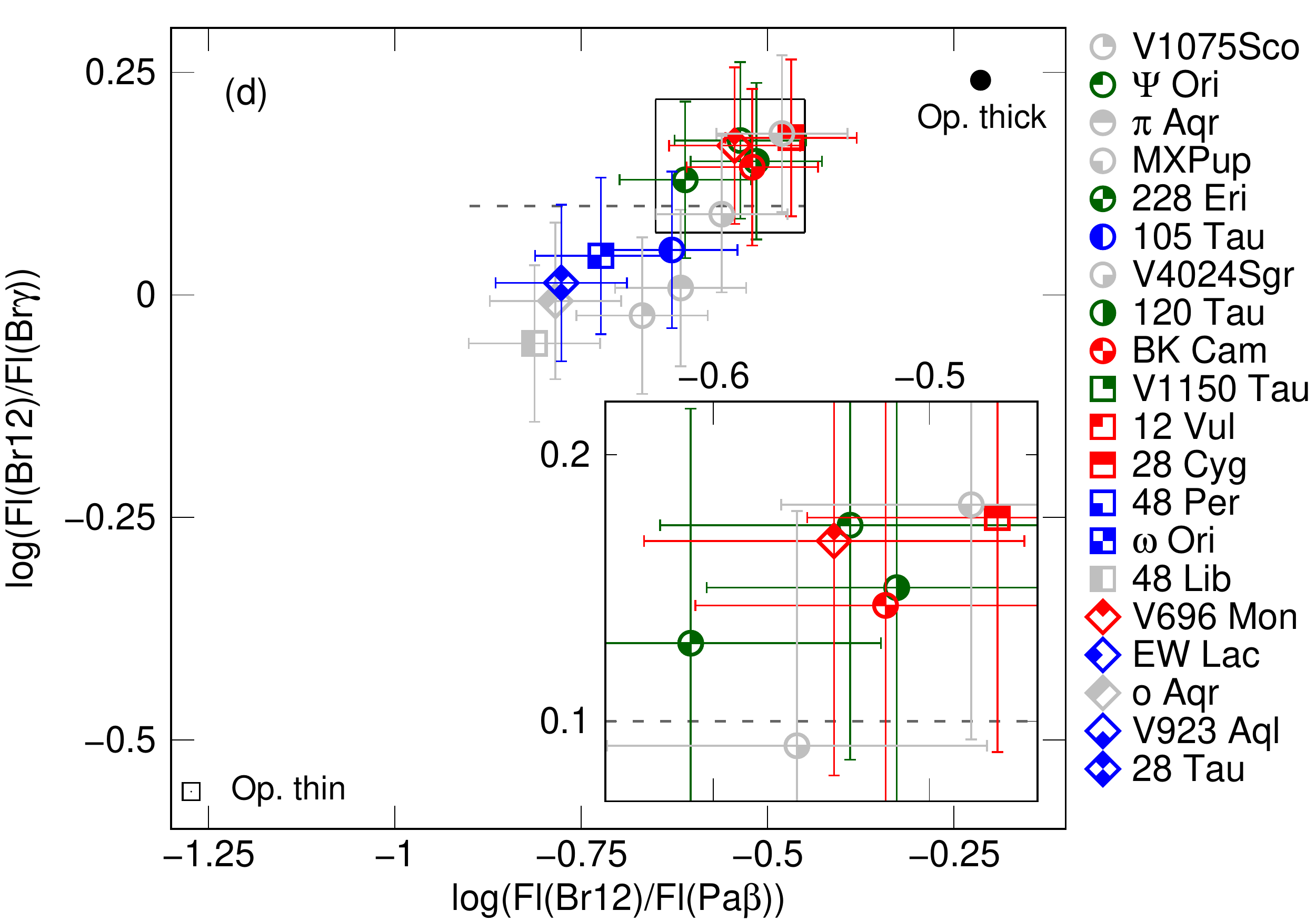}
\caption{Flux-ratio diagrams based on the following lines: (a) \mbox{Hu14 - Pf$\gamma$ - Br$\alpha$} (Lenorzer's); (b) \mbox{Br$\alpha$ - Pf$\gamma$ - Br$\gamma$} (Granada's - Persson \& McGregor's); (c) \mbox{Pf18 - Br$\gamma$ - Pa$\beta$}; (d) \mbox{Br12 - Pa$\beta$ - Br$\gamma$}. The symbols for each star are the same as in Fig.~\ref{fig:EWBraBrg}, but the colour represents the group the star belongs to: red symbols represent group I stars, blue symbols group II stars, and green symbols the transition group I-II. The grey symbols are for those stars without an L-band spectrum and no group classification. Zoomed-in views of the region where symbols overlap are plotted in panels (c) and (d).}
\label{fig:fluxratiodiagrams}
\end{figure*}

For a dense gas, lines are optically thick and their line strengths only depend on the emitting surface. On the contrary, for low gas densities, the lines are optically thin and can be represented by Menzel's case B recombination \citep{BakerMenzel1938}. \citet{Lenorzer2002Diagrama} made a diagram of the flux ratios Hu14/Br$\alpha$ versus Hu14/Pf$\gamma$ and plotted the location of different objects, together with the optically thick and optically thin limits. On this diagram, LBVs, Be, and B[e] stars are located in different loci along the diagonal between the optically thick and optically thin cases: LBVs are located in the lower part, close to the optically thin limit at (\mbox{log(Fl(Hu14)/Fl(Pf$\gamma$))}, \mbox{log(Fl(Hu14)/Fl(Br$\alpha$))})~$\simeq$~\mbox{(-0.9,-1.6)}, Be stars are close to the optically thick case at (\mbox{log(Fl(Hu14)/Fl(Pf$\gamma$))}, \mbox{log(Fl(Hu14)/Fl(Br$\alpha$))})~$\simeq$~\mbox{(0,0)} in the upper part, while B[e] stars are in the central part. Moreover, \citet{Mennickent2009} located group I and II stars in a Lenorzer diagram and found that the group I stars are located in the top right part of the Lenorzer diagram, pointing out a compact and dense disc structure, while group II stars are more spread out and located in regions of moderate-to-small-line optical depths. Thereby, the location of the objects in the diagram gives us information about the density of the emitting gas.

Figure~\ref{fig:fluxratiodiagrams}~a shows the Lenorzer diagram for those programme stars observed in the L band. The colour of the symbols in this figure is the same as in Fig~\ref{fig:EWBraBrg}. Stars of group I (BK\,Cam, 12\,Vul, 28\,Cyg, and V696\,Mon) are located in the region log(Fl(Hu14)/Fl(Br$\alpha$)) $\geq$ -0.25 (limit marked with a solid black line), value that is close to the limit of -0.2 adopted by \citet{Mennickent2009} for stars in this group. For stars with log(Fl(Hu14)/Fl(Br$\alpha$)) below this value, classified as belonging to group II according to the intensity of Humphreys lines, we can distinguish between the ones that present EW(Br$\alpha$)/EW(Br$\gamma$) in the range expected for group II stars, and those with values nearer to the ones expected for stars of group I. Stars with EW(Br$\alpha$)/EW(Br$\gamma$) ratios in the expected range for group II stars (105\,Tau, 48\,Per, $\omega$\,Ori, EW\,Lac, V923\,Aql, and 28\,Tau) are below the limit \mbox{log(Fl(Hu14)/Fl(Br$\alpha$))~=~-0.5} (dashed blue line). The others, which are those that present EW(Br$\alpha$)/EW(Br$\gamma$) ratios similar to those expected for group I ($\Psi_{01}$\,Ori, 228\,Eri, 120\,Tau, and V1150\,Tau), are located in the intermediate region -0.5~$\leq$~log(Fl(Hu14)/Fl(Br$\alpha$))~$\leq$~-0.25. 

\citet{Granada2010} made a diagram similar to that of \citet{PerssonMcGregor1985} based on the Br$\alpha$, Pf$\gamma$, and Br$\gamma$ lines. They reported that these lines are also useful for distinguishing between compact thick or extended thin envelopes. Figure~\ref{fig:fluxratiodiagrams}~b presents this diagram for our star sample, where stars of group I and I-II are located in the left part, while those of group II are located in the right part.

As observations in the L band are more difficult to obtain and both previously shown diagrams use lines from this band, we consider it useful to make a new diagram only based on lines from J, H, and K bands. With this aim, we made the diagrams shown in Figs.~\ref{fig:fluxratiodiagrams}~c and d, which use the \mbox{Pf18 - Br$\gamma$ - Pa$\beta$} and \mbox{Br12 - Pa$\beta$ - Br$\gamma$} lines, respectively. The stars plotted in grey are those without an L-band spectrum and, therefore, without group classification and not plotted in Figs.~\ref{fig:fluxratiodiagrams}~a and b.

The distributions obtained for the diagrams based on \mbox{Pf18 - Br$\gamma$ - Pa$\beta$} and \mbox{Br12 - Pa$\beta$ - Br$\gamma$} lines are similar to the one observed in the Lenorzer diagram. The group II stars are close to the optically thin limit in the lower part of the diagram at (\mbox{log(Fl(Pf18)/Fl(Br$\gamma$))}, \mbox{log(Fl(Pf18)/Fl(Pa$\beta$))})~$\simeq$~\mbox{(-1.4,-2.1)} and (\mbox{log(Fl(Br12)/Fl(Pa$\beta$))}, \mbox{log(Fl(Br12)/Fl(Br$\gamma$))})~$\simeq$~\mbox{(-1.25,-0.6)}, while the group I stars are in the opposite corner, close to the optically thick case located at 
(\mbox{log(Fl(Pf18)/Fl(Br$\gamma$))}, \mbox{log(Fl(Pf18)/Fl(Pa$\beta$))})~$\simeq$~\mbox{(-0.25,-0.6)} and (\mbox{log(Fl(Br12)/Fl(Pa$\beta$))}, \mbox{log(Fl(Br12)/Fl(Br$\gamma$))})~$\simeq$~\mbox{(-0.2,-0.25)}.
The location of stars in group I-II is similar to that of stars in group I, and there is one star (105 Tau) classified as group II that falls in the upper region of the \mbox{Pf18 - Br$\gamma$ - Pa$\beta$} diagram. It is worth mentioning that the position of 105 Tau in the Lenorzer diagram is over the line that divides the group II and I-II regions, so its location in Fig.~\ref{fig:fluxratiodiagrams}~c suggests that it is likely an object belonging to group I-II. This intermediate classification may also indicate that these objects are in a transition stage between groups.

For the stars without group classification, there are three ($\pi$\,Aqr, 48\,Lib and $o$\,Aqr) that are in the middle of the gap that separates the groups in Fig.~\ref{fig:fluxratiodiagrams}~c and are located in a region compatible with group II in Fig.~\ref{fig:fluxratiodiagrams}~d. V1075\,Sco is located in the same region in Fig.~\ref{fig:fluxratiodiagrams}~d. Thus, these four stars are likely part of group II.

Regarding the other two stars, in Fig.~\ref{fig:fluxratiodiagrams}~d MX\,Pup is in a region compatible with group I stars, while V4024\,Sgr falls in an intermediate locus. Based on that, MX\,Pup likely belongs to group I, and we can not assign a group for V4024\,Sgr. 

\subsection{Helium and metallic lines}

We identify lines from He~{\sc i}, C~{\sc i}, N~{\sc i}, O~{\sc i}, Na~{\sc i}, Mg~{\sc i}, Mg~{\sc ii}, Si~{\sc i}, Fe~{\sc i}, and Fe~{\sc ii}. In Tables~\ref{tabla:metales1} to \ref{tabla:metales3}, we present the EWs of the identified lines. As we mentioned before, positive EW indicate emission features, while negative EW indicate absorption features.

He~{\sc i} lines are present in 18 stars of our sample, generally in absorption, with the exception of the He I line at $\lambda\,2.0587\,\mu$m which is in emission. The Fe~{\sc i} and Fe~{\sc ii} lines are observed in emission in 16 and in 13 stars observed in the H band, respectively. In some of the stars that present Fe~{\sc ii} lines in emission in the H band, Fe~{\sc ii} lines are also observed in emission in the K and L bands. For the 19 objects observed in the J band, 15 stars present emission lines of O~{\sc i}, and 14 stars present C~{\sc i} in emission. Both elements also appear in the H- and L bands.
Mg~{\sc ii} doublet $\lambda\lambda2.1375-2.1438\,\mu$m is observed in 11 stars of the sample, while the other doublet in $\lambda\lambda2.4038-2.4131\,\mu$m is more difficult to distinguish, because it is blended with H lines. Mg~{\sc ii} is also observed in the H band.
In the H band, only five objects present N~{\sc i} emission at $\lambda\,1.5586\,\mu$m, while four present emission from Si~{\sc i} $\lambda\,1.5964\,\mu$m and $\lambda\,1.6064\,\mu$m.
Two objects show Na~{\sc i} $\lambda\,2.2062\,\mu$m and $\lambda\,2.2090\,\mu$m in emission in the K band. 

\section{Discussion}\label{discussion}

\subsection{An improved criterion to classify Be stars according to the disc opacity}\label{criteria}

\citet{Mennickent2009}'s criterion provides an easy way to separate stars in groups I and II through a qualitative description of the L-band spectrum of the Be stars, and thus to obtain information about the density of the emitting gas. However, in some cases where the Humphreys lines are quite intense, but not as intense as the Br$\alpha$ and Pf$\gamma$ lines, the classification may be ambiguous. In fact, we see that some stars classified as group II objects based on this criterion, actually have EWs and line-flux ratios similar to the ones of group I stars. Thus, we propose that the stars in our group I-II are indeed part of group I. 

Assuming that stars in group I-II are actually part of group I, we can then quantitatively define different criteria that separate stars between groups I and II. These criteria, with limits represented by a dashed grey line in each panel of Fig.~\ref{fig:fluxratiodiagrams}, are shown in Table~\ref{table:criterios}. The criteria were empirically defined based on how the stars of each group gather in the different diagrams, separating group I objects from most group II objects.  In Fig.~\ref{fig:fluxratiodiagrams}~c, the unclassified objects laid in the gap between the group I and group II regions. Therefore, we used the information provided by Fig.~\ref{fig:fluxratiodiagrams}~d, where the same unclassified objects are in positions compatible with group II membership. In Fig.~\ref{fig:fluxratiodiagrams}~d, the limit could be a little displaced, because it is the only diagram where we can plot the unclassified object that is in the gap (V4024\,Sgr).

With these new criteria, we can classify five of the seven stars without L-band observations. MX\,Pup is member of group I, while V1075\,Sco, $\pi$\,Aqr, 48\,Lib, and $o$\,Aqr are members of group II. We can also classify HD\,171623 as a member of group III due to the absence of notorious emission. V4024\,Sgr remains without classification because of its intermediate location in the \mbox{Br12 - Pa$\beta$ - Br$\gamma$} area.

The final group classification for our sample is the following. Group I includes $\psi_{01}$ Ori, MX\,Pup, 228\,Eri, 120\,Tau, BK\,Cam, V1150\,Tau, 12\,Vul, 28\,Cyg, and V696\,Mon; Group II includes V1075\,Sco, $\pi$\,Aqr, 105\,Tau, 48\,Per, $\omega$\,Ori, 48\,Lib, EW\,Lac, $o$\,Aqr, V923\,Aql, and 28\,Tau; Group III includes 66\,Oph and HD\,171623. V4024\,Sgr remains unclassified.

It is worth highlighting that the nine stars that belong to group I have T$_{\mathrm{eff}}\,\geq\,18000$, while seven of the ten stars in group II have T$_{\mathrm{eff}}\,\leq\,18000$. The three stars in group II with T$_{\mathrm{eff}}\,\geq\,18000$ are 105\,Tau, for which we mentioned its limit position in the Lenorzer's diagram, $\pi$\,Aqr and V1075\,Sco. This agrees with the result achieved by \citet{Sabogal2017}, who mentioned that late Be stars present more tenuous and optically thin discs than early-type Be stars, but some early-type stars could present optically thin discs when they lose their discs.

\begin{table}[t!]
\caption{Quantitative criteria to classify stars in groups I and II, according to limits set on Fig.~\ref{fig:fluxratiodiagrams}.} \label{table:criterios}
\centering
\begin{tabular}{l l r} 
\hline\hline
Ratio & Group I & Group II \\
\hline
log(Fl(Hu14)/Fl(Br$\alpha$))                  & $\gtrsim -0.5$  & $\lesssim -0.5$ \\
2.5$\cdot$log(Fl(Br$\alpha$)/Fl(Pf$\gamma$))  & $\lesssim 0.55$ & $\gtrsim 0.55$ \\
log(Fl(Pf18)/Fl(Pa$\beta$))                   & $\gtrsim -1$    & $\lesssim -1$   \\
log(Fl(Br12)/Fl(Br$\gamma$))                  & $\gtrsim 0.1$   & $\lesssim 0.1$  \\
\hline
\end{tabular}
\end{table}

\subsection{Behaviour of the He and metallic lines}

\begin{figure} 
\centering
\includegraphics[angle=270,width=0.45\columnwidth]{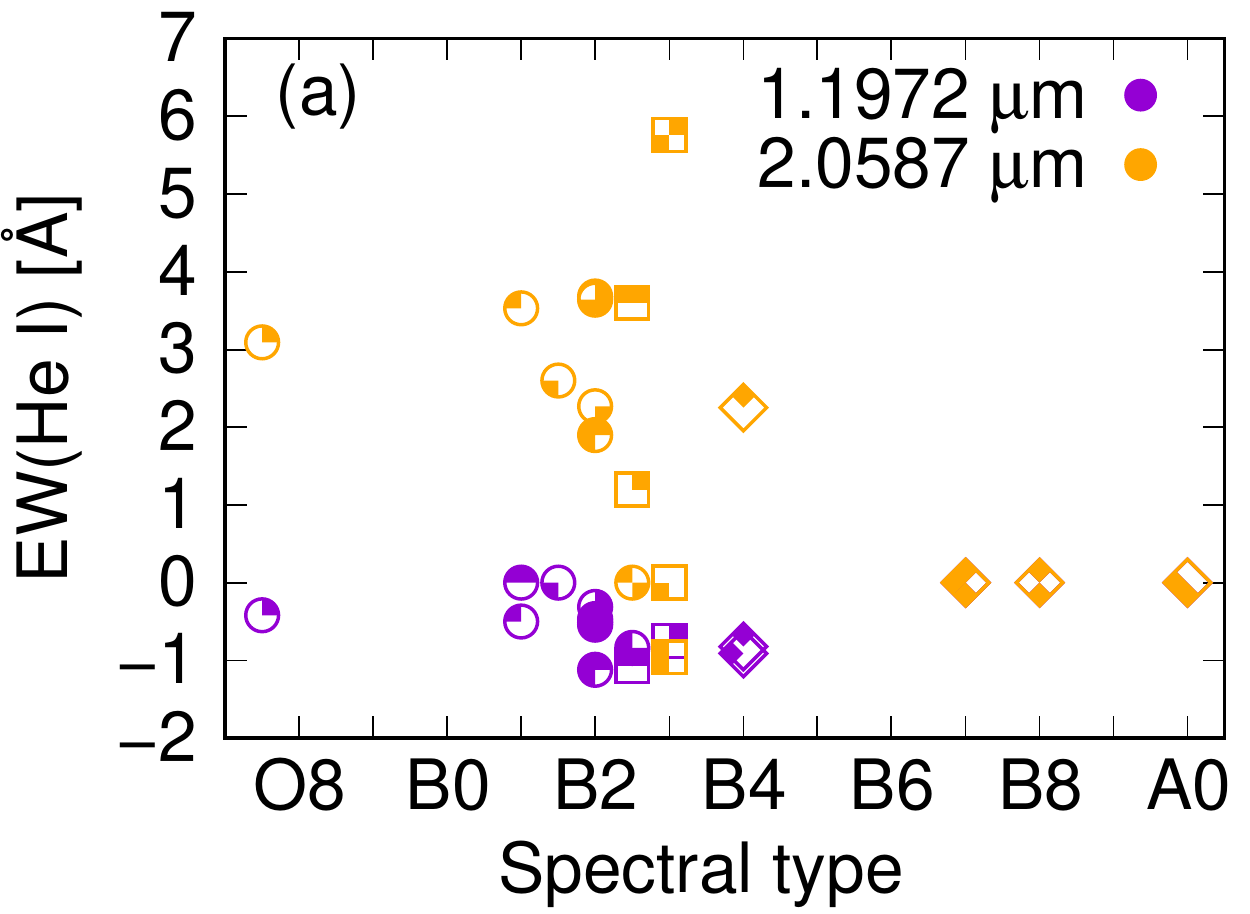}
\includegraphics[angle=270,width=0.45\columnwidth]{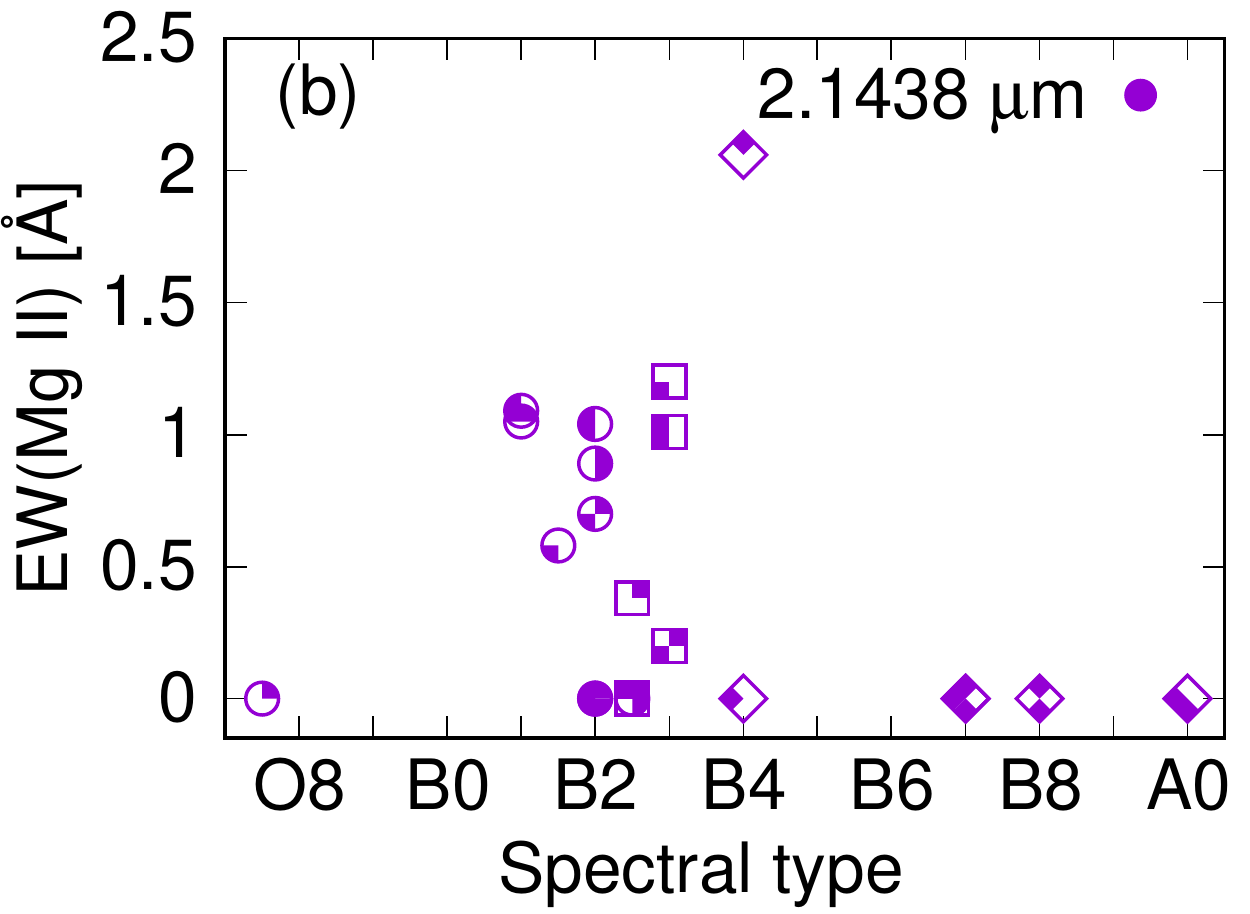}\\
\includegraphics[angle=270,width=0.45\columnwidth]{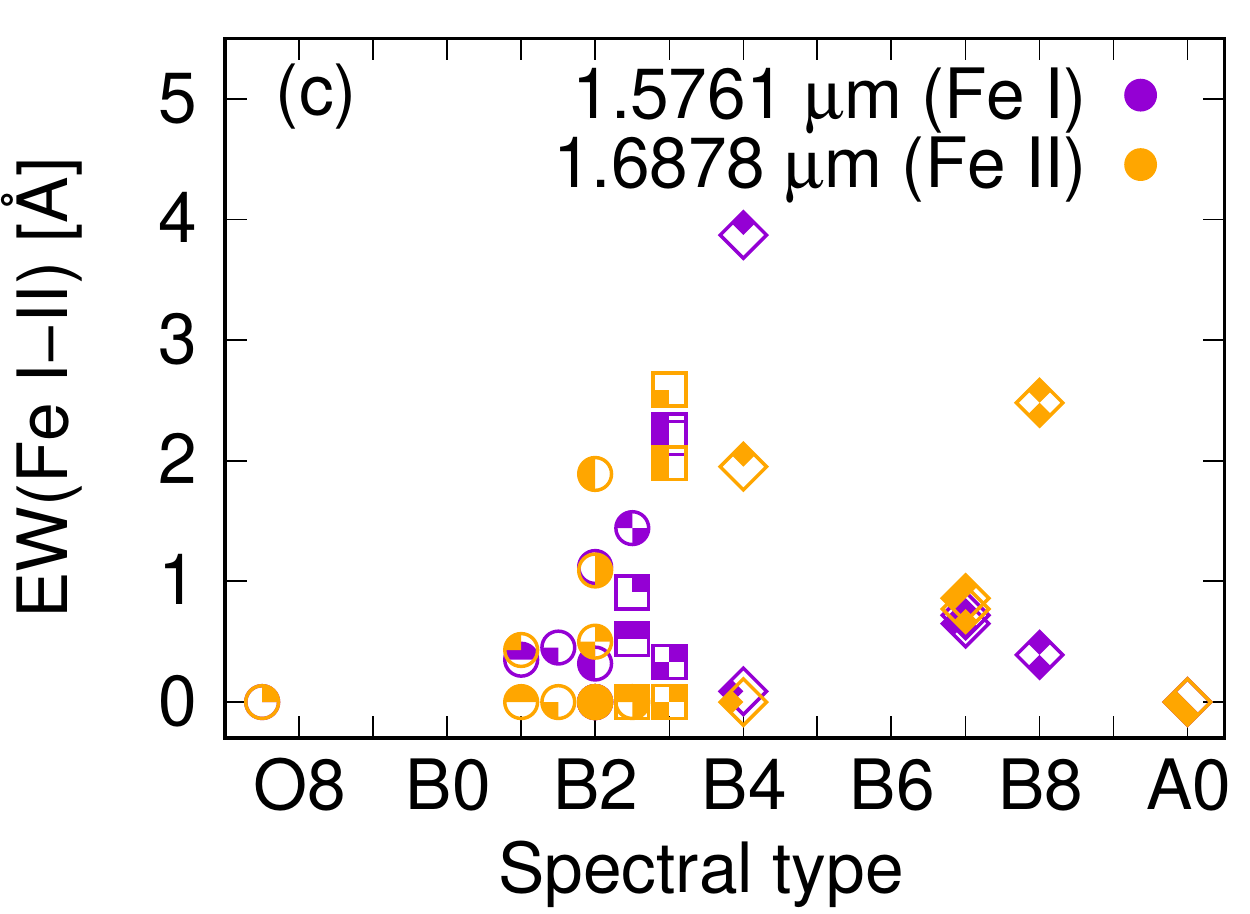}
\includegraphics[angle=270,width=0.45\columnwidth]{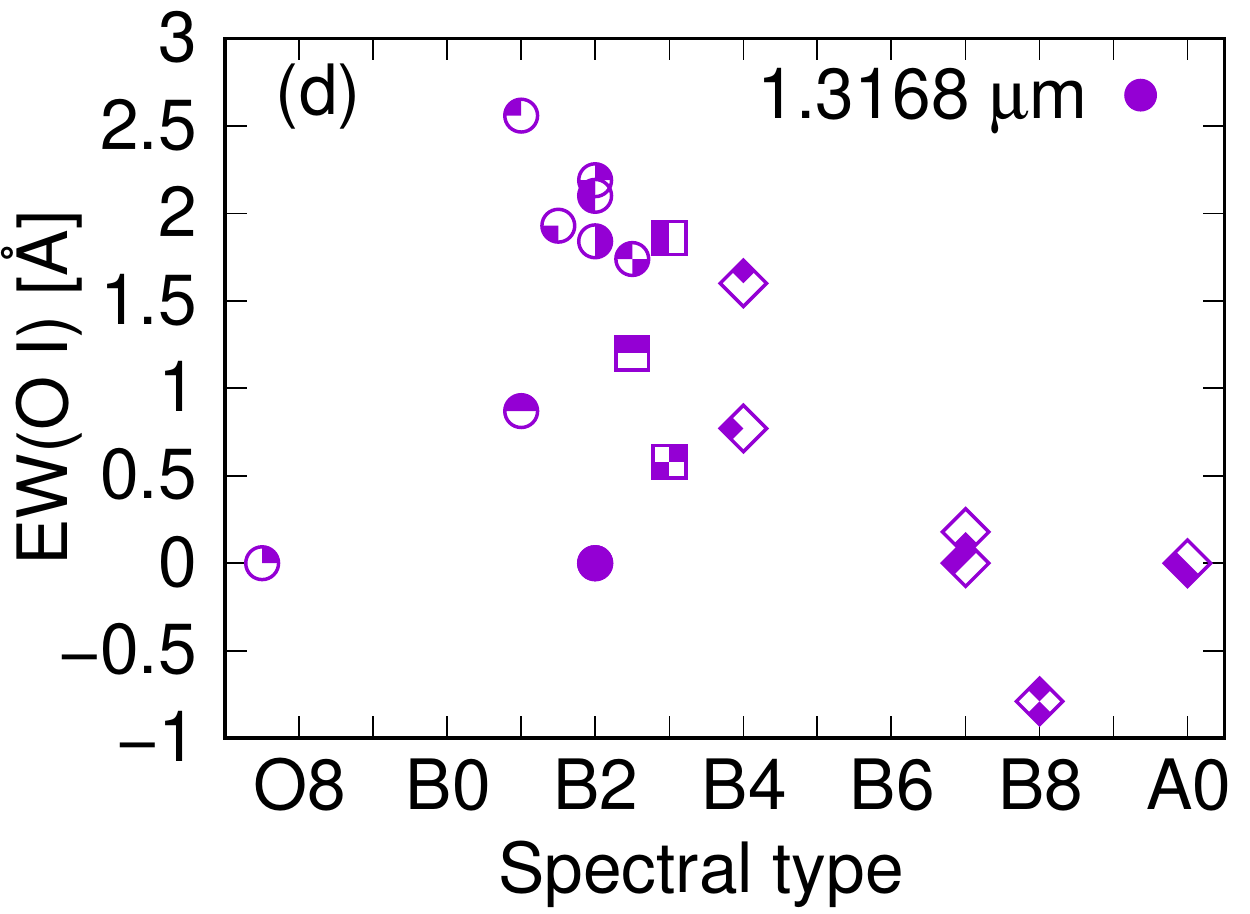}\\
\caption{EW versus ST for representative lines of He~{\sc i}, Mg~{\sc ii}, Fe~{\sc i}, Fe~{\sc ii} and O~{\sc i}. The uncertainties for the EWs are around 10\% of the values.}
\label{fig:TE_metals}
\end{figure}

Figure~\ref{fig:TE_metals} shows the EWs of representative lines from different elements versus the spectral type of the star. The symbols are the same as in Fig.~\ref{fig:EWBraBrg}, but different colours represent different lines of the same element, as is indicated in each panel. 

In Fig.~\ref{fig:TE_metals} a, we plot the EW values from the He~{\sc i}~$\lambda\,2.0587\,\mu$m line in emission and the He~{\sc i}~$\lambda\,1.1972\,\mu$m line in absorption. The He~{\sc i} is observed only in early B-type stars, with EW\,Lac (ST B4) and V696 Mon (ST B3-5) being the stars with the latest spectral type that present this element. This is in good agreement with previous results by \citet{Hanson1996}, \citet{Clark2000} and \citet{Lenorzer2002Atlas}.

The Mg~{\sc ii}~$\lambda\,2.1438\,\mu$m line is only observed in the same spectral types as He lines (see Fig.~\ref{fig:TE_metals} b). This agrees with the results by \citet{Clark2000}, who said that Mg II lines are likely excited via Ly$\beta$ fluorescence \citep{Bowen1947} and are observed in stars with spectral types earlier than B4.

Regarding Fe~{\sc i}~$\lambda\,1.5761\,\mu$m and Fe~{\sc ii}~$\lambda\,1.6878\,\mu$m lines (Fig.~\ref{fig:TE_metals} c), it seems that the emission is maximum for spectral types B2-B4. Through the comparison between the EWs of Fe~{\sc ii}, Mg~{\sc ii} and Br$\gamma$, \citet{Clark2000} proposed that neither Fe~{\sc ii} nor Mg~{\sc ii} lines are seen in emission when EW(Br$\gamma$)$<8\,\AA$.
We only have one object with EW(Br$\gamma$)$<8\,\AA$ (EW\,Lac), which does not present Fe~{\sc ii} or Mg~{\sc ii} emission in the K band.

\citet{Lenorzer2002Atlas} reported that O~{\sc i} emission lines in the L band are observed in Be stars earlier than B3. However, in the K band we observe O~{\sc i}~$\lambda\,1.3168\,\mu$m emission lines in stars with spectral types up to B7. It is worth mentioning that, as is shown in Fig.~\ref{fig:TE_metals} d, the EW of O~{\sc i} emission lines becomes smaller for later spectral-type objects. This relation could be useful to estimate the ST of the star. In all the cases, EW(1.3168$\,\mu$m)/EW(1.129$\,\mu$m)<1, as is expected if the Ly$\beta$ fluorescence process has a significant role in the excitation of the O~{\sc i} lines. 

For the C~{\sc i}~$\lambda\,1.1756\,\mu$m line, the existence of a correlation is not clear. \citet{Clark2000} mentioned that the presence of Na~{\sc i} emission features suggests that a region of the circumstellar envelope should be shielded from direct stellar radiation, due to the low ionisation energy. This could be the case for 48\,Per and V696\,Mon, which are the two objects of our sample that present emission in Na~{\sc i} lines in the K band.

\subsection{Correlation with $V\,\sin\,i$}

To discuss if the main mechanism of broadening is rotational, we analyse the correlation between the FWHM and $\Delta \mathrm{V}$ parameters versus $V\,\sin\,i$. A strong correlation between the FWHM and $V\,\sin\,i$ was found by \citet{Hanuschik1989} for optical lines and \citet{Granada2010} for IR lines. 

The determination of the FWHM and $\Delta \mathrm{V}$ depends on the spectral resolution; therefore, to analyse their correlation with $V\,\sin\,i$ we selected GNIRS spectra. These spectra have the same resolution, and the sample is bigger than that observed with FIRE. 

\begin{figure} 
\centering
\includegraphics[angle=270,width=0.45\columnwidth]{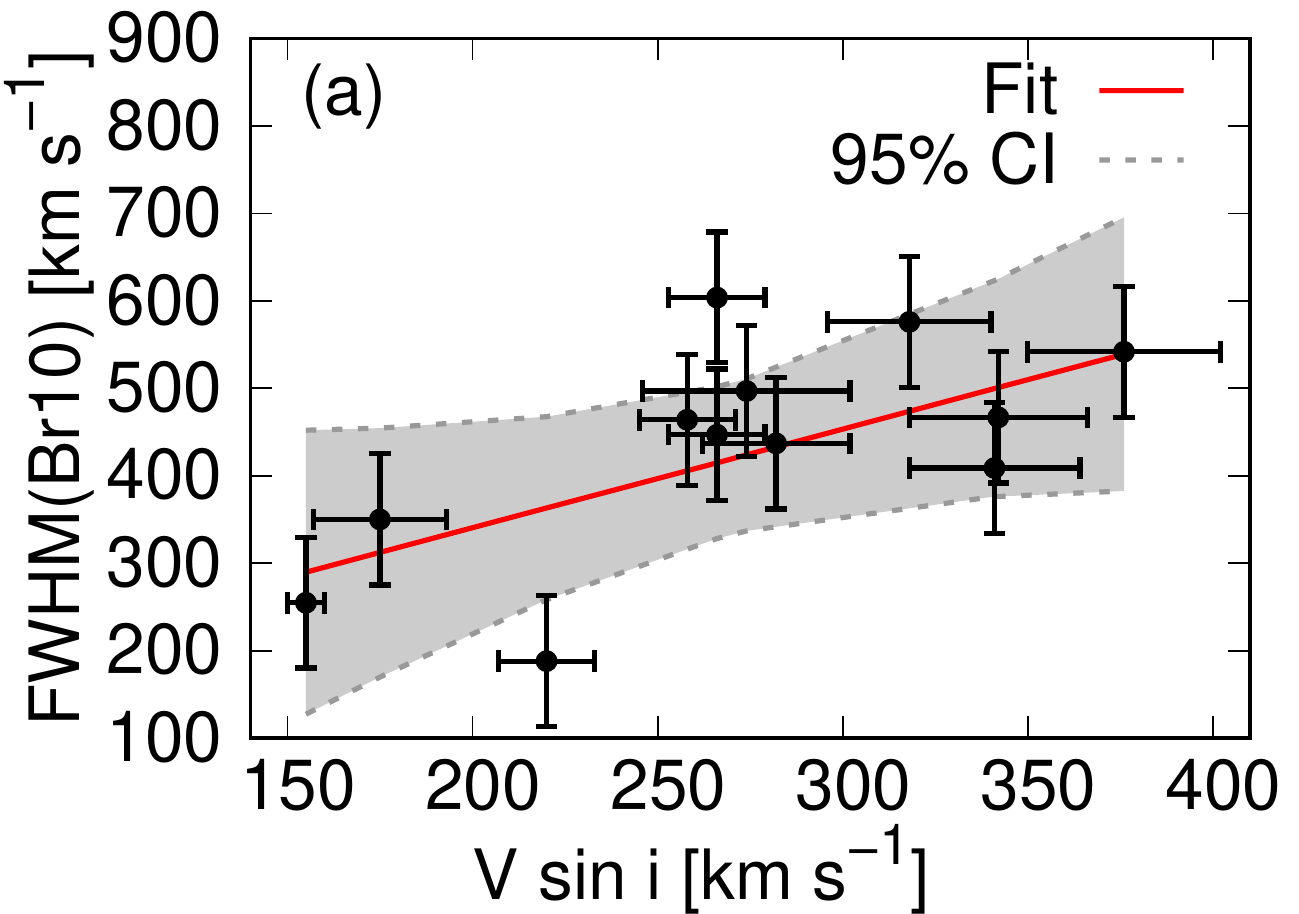}
\includegraphics[angle=270,width=0.45\columnwidth]{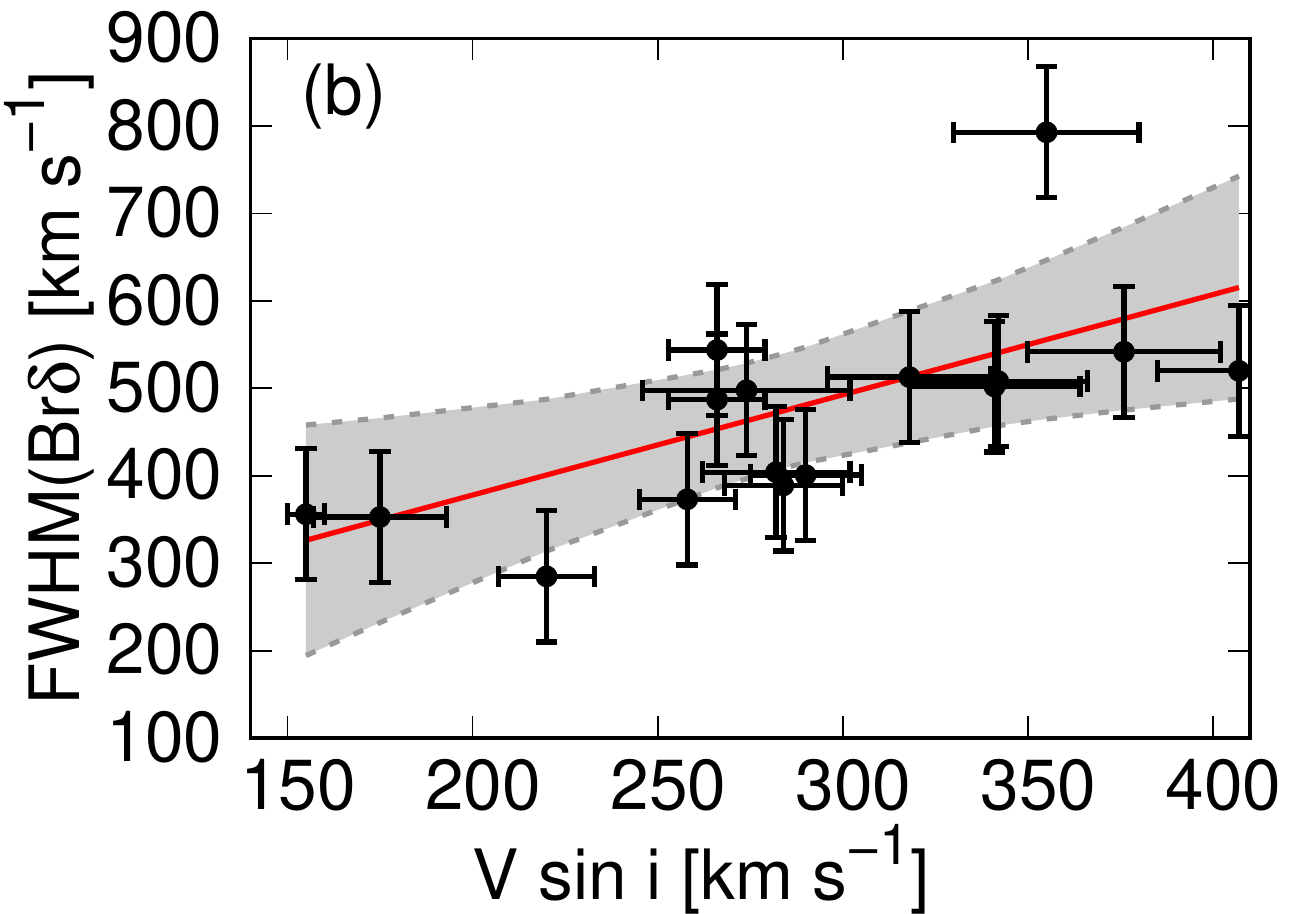}
\includegraphics[angle=270,width=0.45\columnwidth]{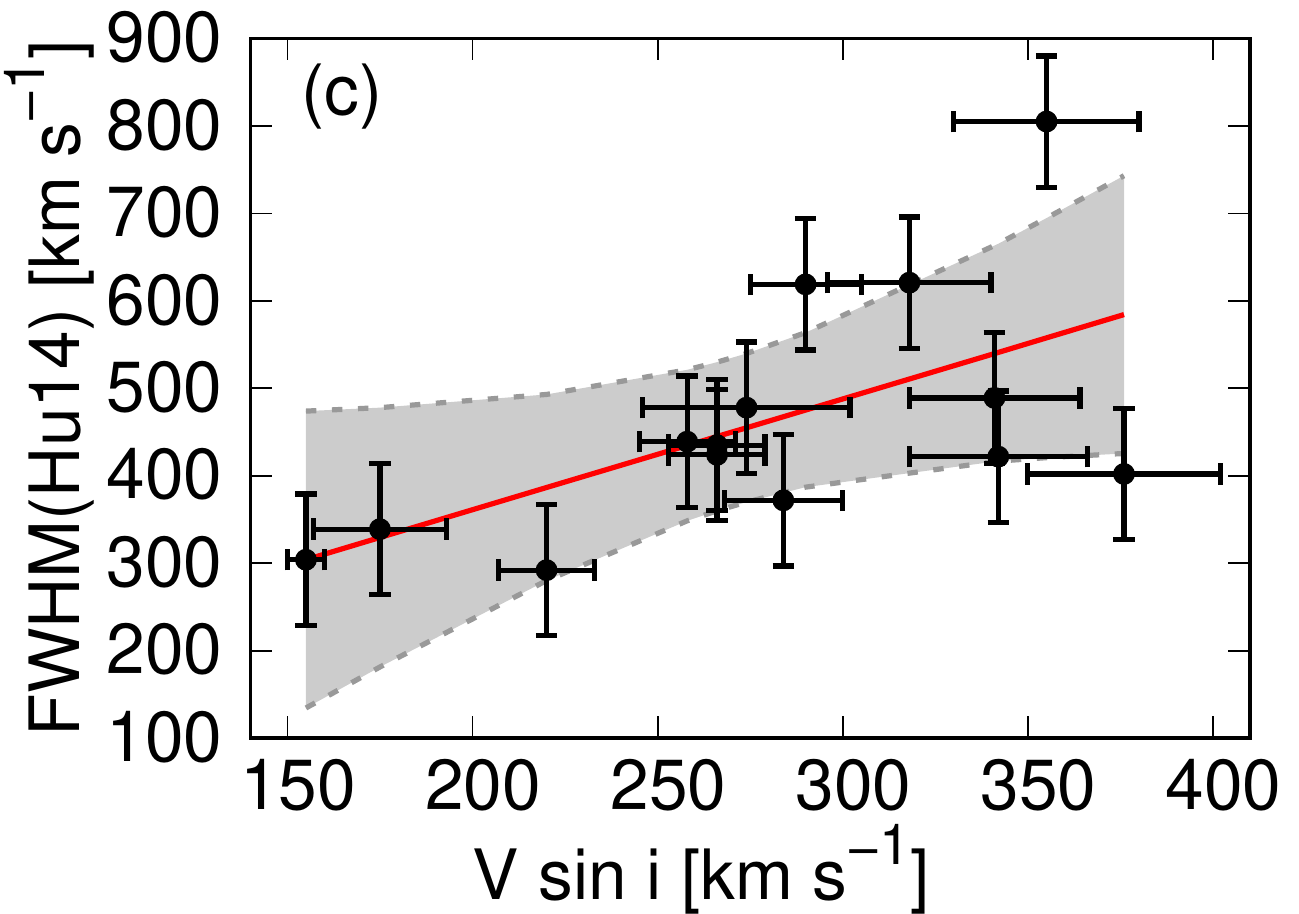}
\includegraphics[angle=270,width=0.45\columnwidth]{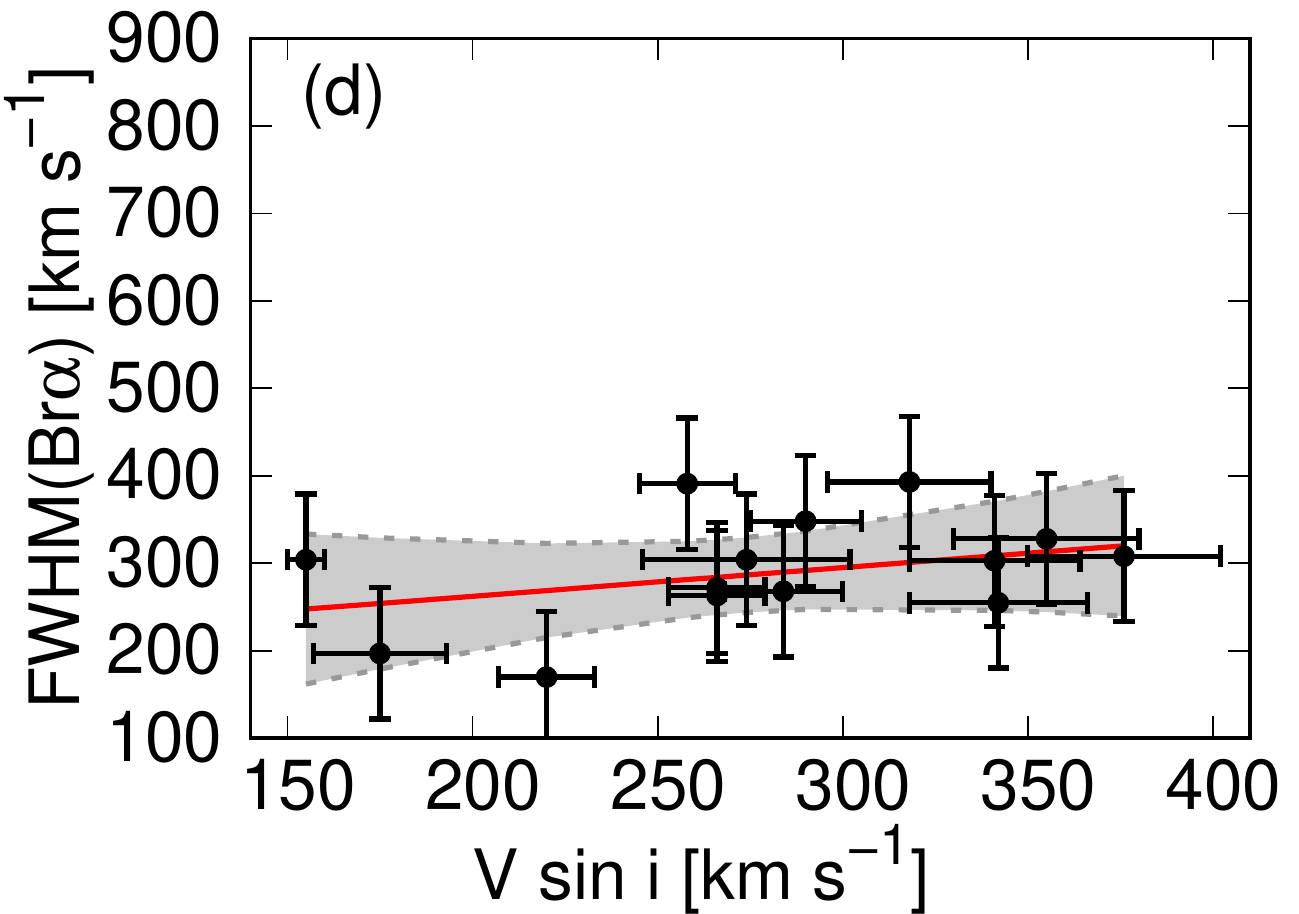}
\caption{FWHMs of Br10 (a), Br$\delta$ (b), Hu14 (c), and Br$\alpha$ (d) lines versus $V\,\sin\,i$. Linear fits (red line) and 95\% confidence intervals (grey area) are superimposed in each panel. }
\label{fig:Vsini_FWHM}
\end{figure}

In Fig.~\ref{fig:Vsini_FWHM}, we plot the FWHM values of the Br10, Br$\delta$, Hu14, and Br$\alpha$ lines versus $V\,\sin\,i$. In each panel, a linear fitting and the 95\% confidence interval are superimposed with a red line and grey area. To quantify the statistical relationship between the FWHM and $V\,\sin\,i$, we calculated the Pearson correlation coefficient for the distributions. For Br10, Br$\delta$, and Hu14 lines (Figs.~\ref{fig:Vsini_FWHM} a, b, and c) we found a strong correlation with Pearson's coefficients of 0.65, 0.67, and 0.60, respectively. Contrary to \citet{Granada2010}, we found no correlation between the FWHM versus $V\,\sin\,i$ for the lines Pf$\gamma$ and Br$\alpha$ (see Fig.~\ref{fig:Vsini_FWHM} d), nor for Br$\gamma$, Pa $\alpha$, Pa$\beta$, or Pf$\delta$ lines.

\begin{figure} 
\centering
\includegraphics[angle=270,width=0.45\columnwidth]{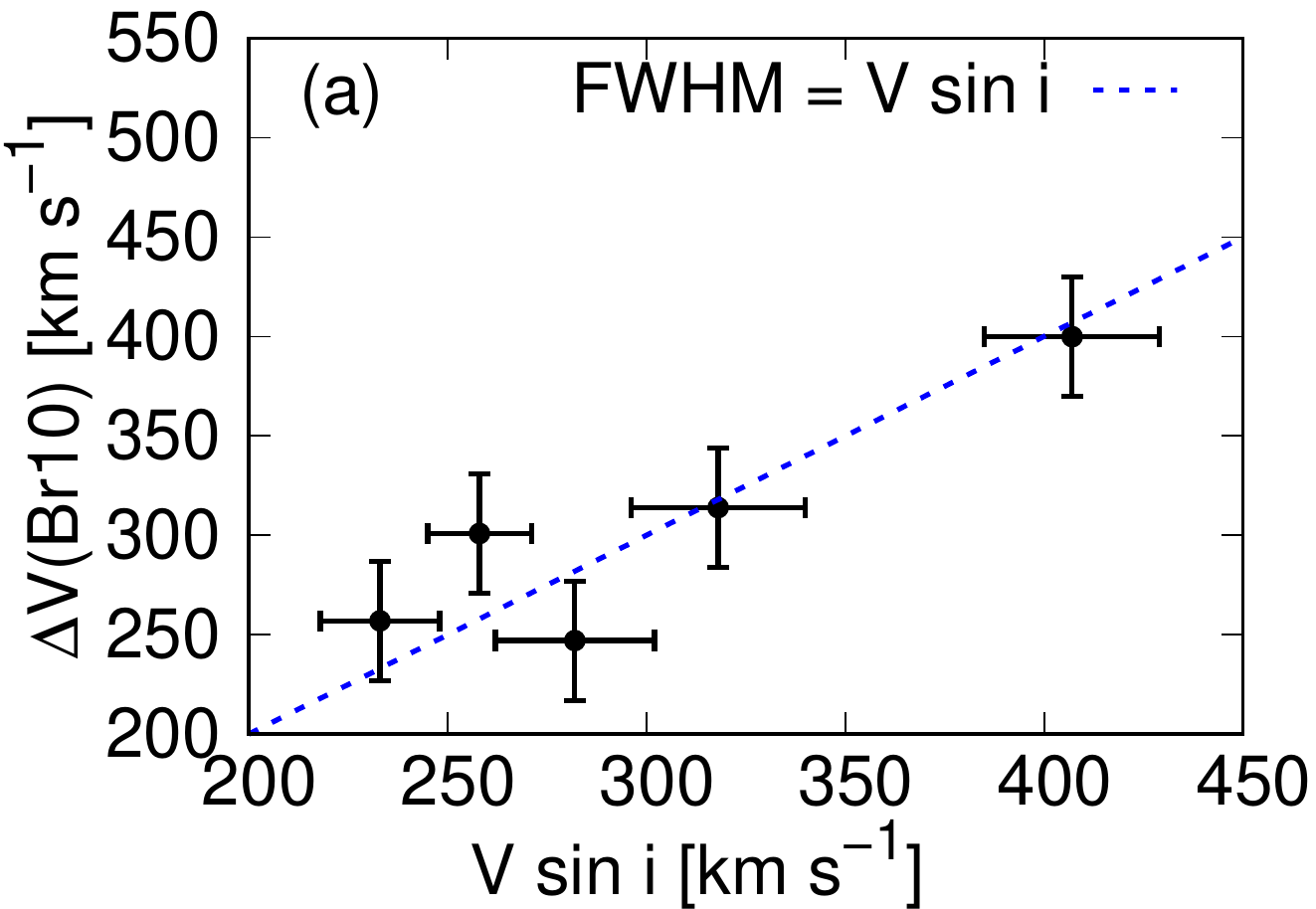}
\includegraphics[angle=270,width=0.45\columnwidth]{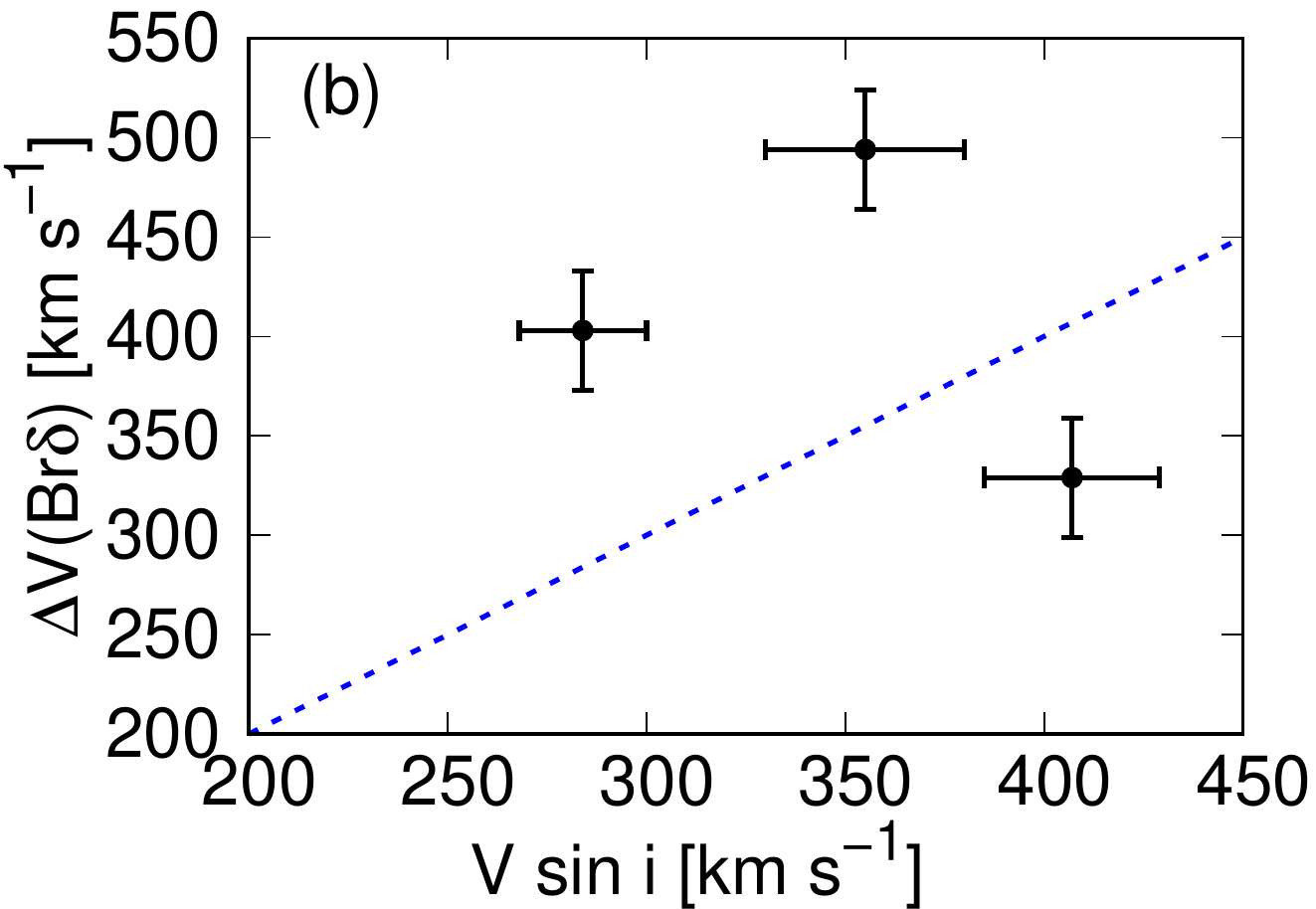}
\includegraphics[angle=270,width=0.45\columnwidth]{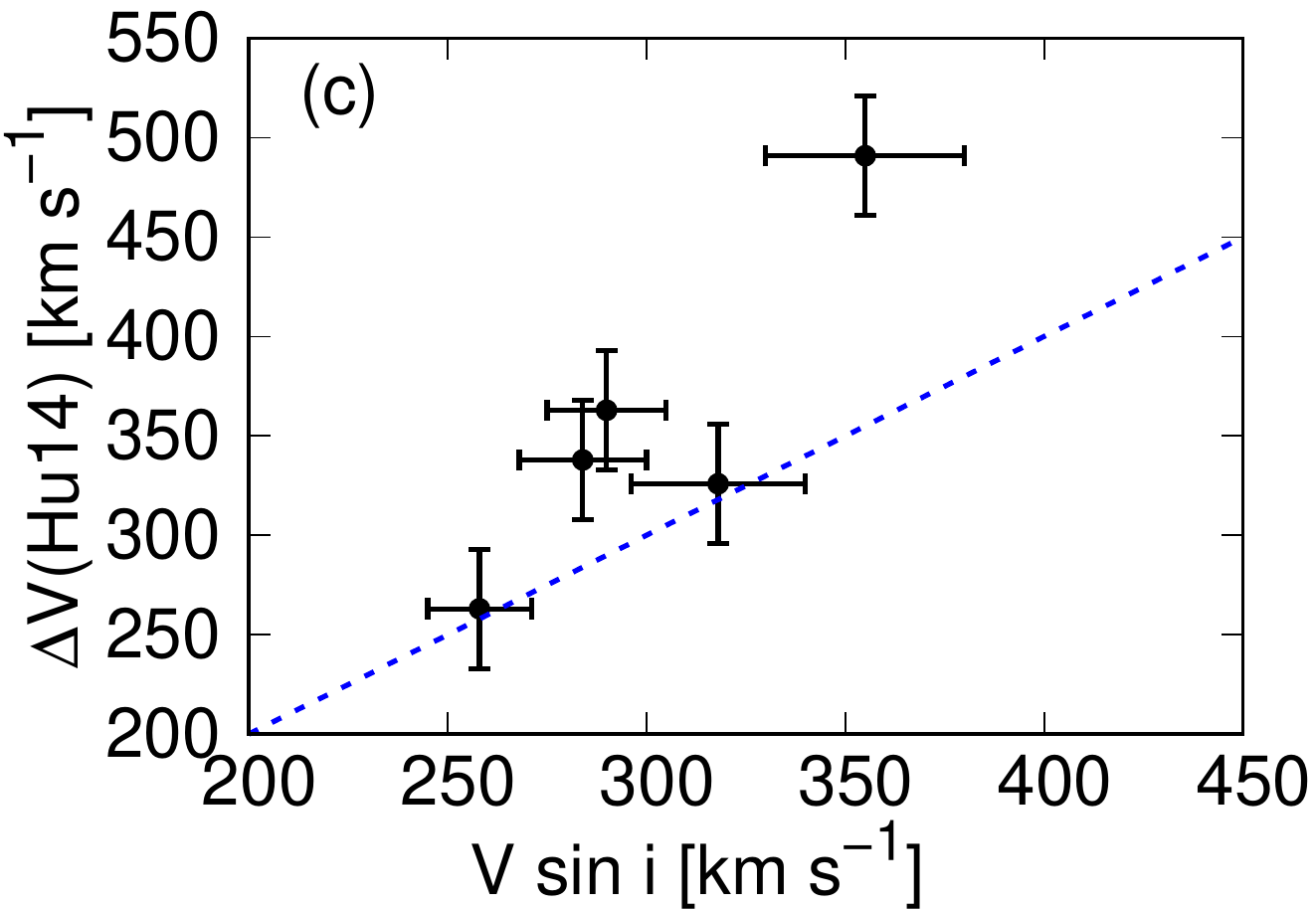}
\caption{Peak separation of the Br10 (a), Br$\delta$ (b) and Hu14 (c) lines versus $V\,\sin\,i$. The linear relation $\Delta \mathrm{V} = V\,\sin\,i$ is represented by a dashed blue line. }
\label{fig:Vsini_DV}
\end{figure}

Figure~\ref{fig:Vsini_DV} presents the peak separation versus $V\,\sin\,i$ for the same lines that the FWHM correlates with $V\,\sin\,i$ (Br10, Br$\delta$ and Hu14). It is seen that the $\Delta \mathrm{V}$ values on the Br10 line follow the relationship $\Delta \mathrm{V}= V\,\sin\,i$, which is plotted via a in dashed blue line. For the Hu14 line, the correlation is weaker, and there is no correlation for the Br$\delta$ line. 

It is worth mentioning that those FWHMs and $\Delta \mathrm{V}$ that better correlate to $V\,\sin\,i$ correspond to higher-order Humphreys and Brackett lines, which form in inner parts of the disc compared the low-order lines \citep{Hony2000,Mennickent2009}, which form in outer parts. So, these lines might be useful when estimating the $V\,\sin\,i$ of the stars.

\section{Conclusions}\label{conclusions}

The IR spectra of Be stars are excellent sources of information on the physical and dynamical structure of different regions inside the disc. Their analysis contributes to a better understanding of the possible mechanism involved in the development and evolution of the disc. However, a great number of NIR spectra and a wide spectral range coverage are required to attain a more complete overview of the Be phenomenon. 

With this aim, we present an atlas of quasi simultaneous J-, H-, K-, and L-band observations of a sample of 22 Be stars obtained with the GNIRS and FIRE spectrographs. We identify several lines of the Paschen, Brackett, Pfund and Humphreys series of hydrogen. Lines of He~{\sc i}, C~{\sc i}, N~{\sc i}, O~{\sc i}, Na~{\sc i}, Mg~{\sc i}, Mg~{\sc ii}, Si~{\sc i}, Fe~{\sc i}, and Fe~{\sc ii} are also identified. 

He~{\sc i} lines are observed in early B-type stars. In the same spectral types, the Mg~{\sc ii} and O~{\sc i} lines are observed in emission. Moreover, we report that the intensity of the O~{\sc i}~$\lambda\,1.3168\,\mu$m emission line decreases towards the later spectral types. This reveals the importance of the Ly$\beta$ fluorescence process on Mg~{\sc ii} and O~{\sc i} transitions. 

We also find a correlation between the FWHM of the Br10, Br$\delta$, and Hu14 lines with $V\,\sin\,i$. This indicates that the broadening mechanism is mainly rotational. In addition, the peak separation of Br10 line correlates strongly with $V\,\sin\,i$, while the correlation is weaker for the Hu14 line. We thus propose that these lines are useful when estimating the projected rotational velocity of Be stars.

Through the analysis of different hydrogen line-flux ratio diagrams, and taking into account the different value ranges from the EW(Br$\alpha$)/EW(Br$\gamma$) ratio, we propose new criteria to classify the Be stars into Mennickent's groups. These new criteria also allow us to unify the classification that comes from previous works, which are the definition of the groups given by \citet{Mennickent2009} and the location of the different groups on Lenorzer's diagram reported by \citet{Granada2010}. Moreover, the new criteria involve lines not only in the L band, which are more difficult to obtain, but also in the J, H, and K bands. Using these criteria, we can identify some objects that present compact thick envelopes in our sample (group I stars: $\psi_{01}$ Ori, MX\,Pup, 228\,Eri, 120\,Tau, BK\,Cam, V1150\,Tau, 12\,Vul, 28\,Cyg and V696\,Mon), while in others the envelope is extended and optically thin (group II stars: V1075\,Sco, $\pi$\,Aqr, 105\,Tau, 48\,Per, $\omega$\,Ori, 48\,Lib, EW\,Lac, $o$\,Aqr, V923\,Aql and 28\,Tau). The presence of an optically thin or thick disc seems to depend on $T_{\text{eff}}$, since group I stars have $T_{\text{eff}}\geq18000$ while most group II stars have $T_{\text{eff}}\leq18000$. We also find two objects that do not present an emitting disc at the observing time (66\,Oph and HD\,171623), and one star (V4024\,Sgr) remains unclassified.

Paper II of this series will use the line parameters from this work to obtain a characterisation of the envelopes. It could be valuable to discuss the evolution of the envelopes through the analysis of the variability of their physical conditions and to identify some peculiar objects such as 12\,Vul, which has presented a transitory $^{12}$CO emission \citep{Cochetti2021}. A better knowledge of the properties and evolution of the envelopes enable us to set constraints on the theoretical models of the stars with the Be phenomenon. 

\begin{acknowledgements}

Based on observations obtained at the international Gemini Observatory under programs GN-2010B-Q-2, GN-2012B-Q-56, GN-2016B-Q-83, GN-2017A-Q-84, GN-2017A-Q-89, GN-2017B-Q-81, and GN-2017B-Q-86. 

This work has made use of the BeSS database, operated at LESIA, Observatoire de Meudon, France: http://basebe.obspm.fr.

Y.R.C. thank Gabriel Ferrero and Nidia Morrell for allowing her to obtain data in Las Campanas Observatory during the nice stay shared, and acknowledges the support from the Carnegie Institution for Science and Richard Lounsbery Foundation that enable the stay in La Serena and Las Campanas Observatory under the visiting fellowship program for young Argentinian astronomers.

Y.R.C. acknowledges support from a CONICET fellowship.

M.L.A. and A.F.T. acknowledge financial support from the Universidad Nacional de La Plata (Programa de Incentivos 11/G160).

L.S.C. thanks financial support from CONICET (PIP 0177) and from the Agencia Nacional de Promoción Científica y Tecnológica de Argentina (Préstamo BID PICT 2016-1971)

A.G. acknowledges financial support from the Agencia Nacional de Promoción Científica y Tecnológica de Argentina (PICT 2017-3790).

\end{acknowledgements}

   \bibliographystyle{../../../paquetes/aa-package/bibtex/aa} 
   \bibliography{biblio.bib} 

\begin{appendix} 

\section{Spectra of the star sample}\label{app:spectra}

          \begin{figure}[h!]
          \centering
          \includegraphics[angle=0,width=0.94\textwidth]{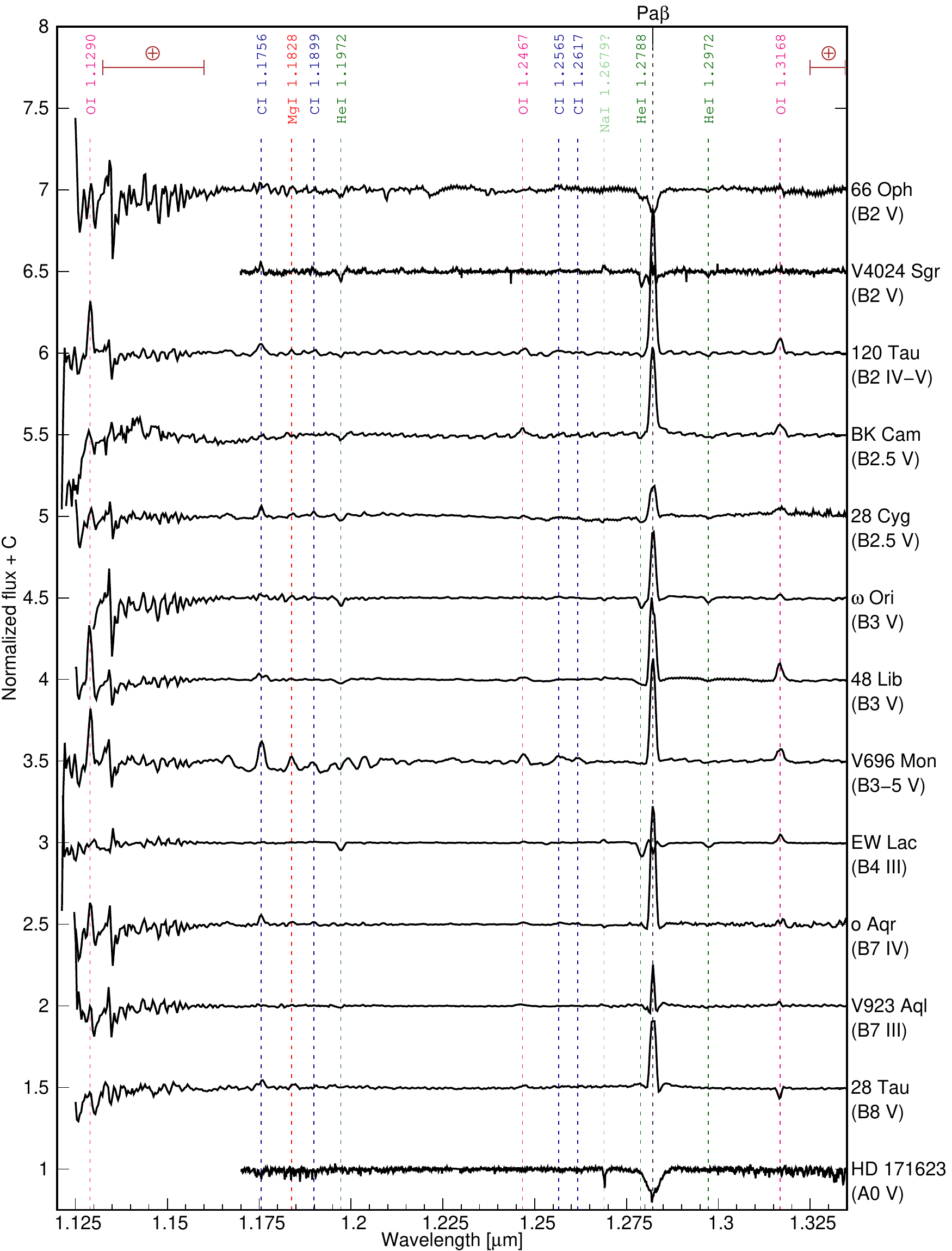}
          \caption{J-Band spectra of Be stars. Spectra are normalised and shifted vertically to ease comparison.}
          \label{fig:AtlasJ2}
          \end{figure}
          \clearpage

          \begin{figure}[b!]
          \centering
          \includegraphics[angle=0,width=0.94\textwidth]{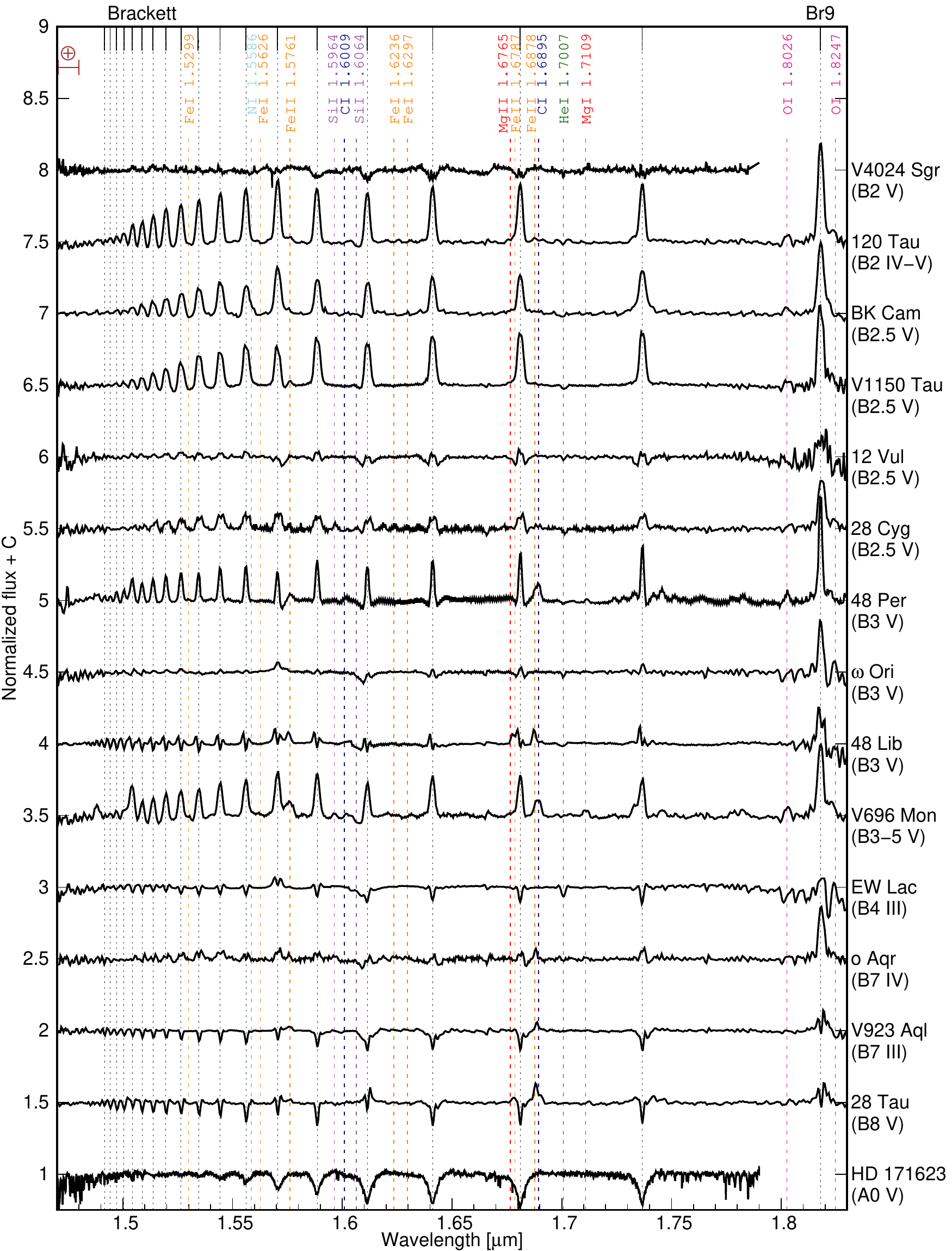}
          \caption{H-band spectra of Be stars. Spectra are normalised and shifted vertically to ease comparison.}
          \label{fig:AtlasH2}
          \end{figure}
          \clearpage

          \begin{figure}[b!]
          \centering
          \includegraphics[angle=0,width=0.95\textwidth]{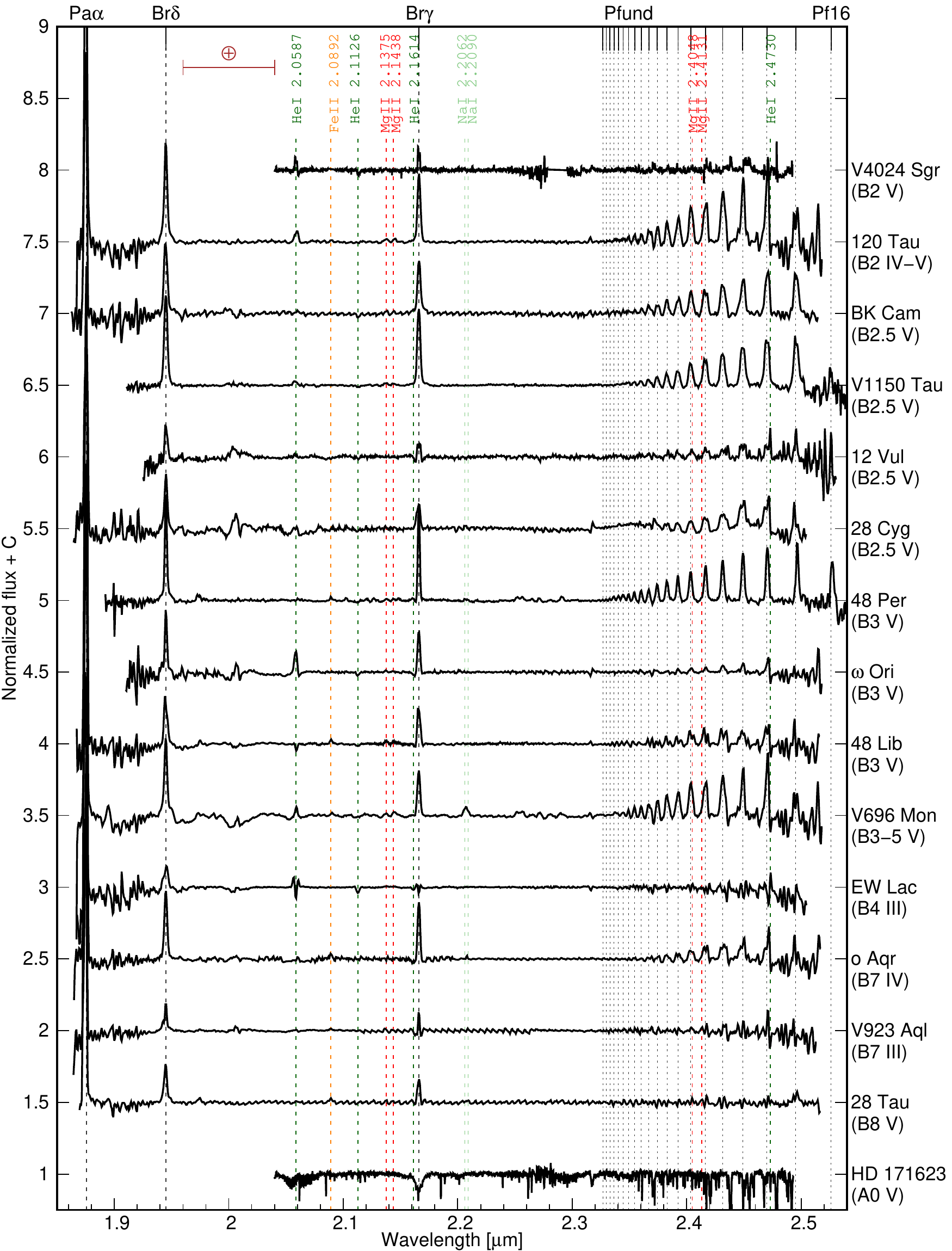}
          \caption{K-band spectra of Be stars. Spectra are normalised and shifted vertically to ease comparison.}
          \label{fig:AtlasK2}
          \end{figure}
          \clearpage

          \begin{figure}[b!]
          \centering
          \includegraphics[angle=0,width=0.95\textwidth]{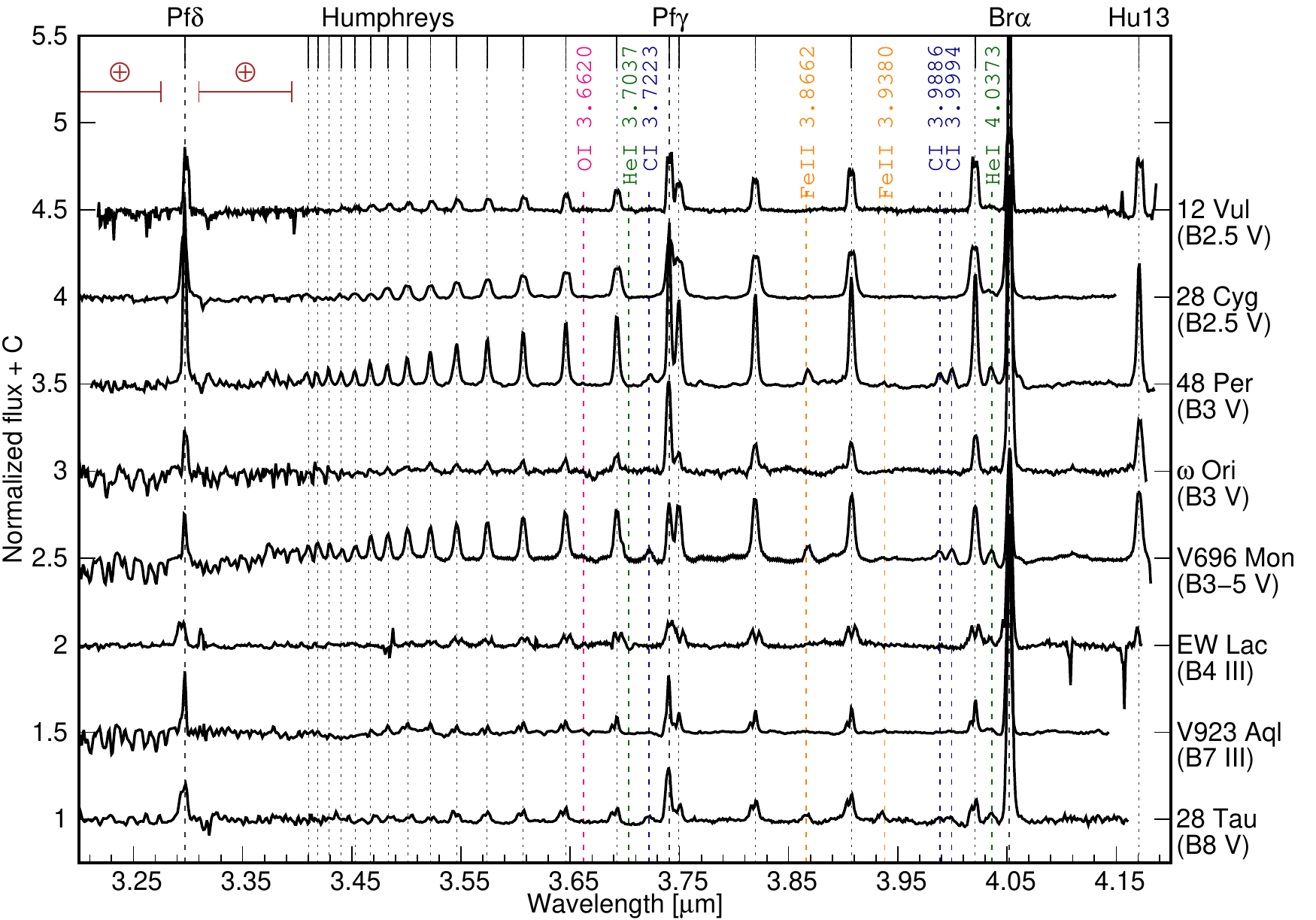}
          \caption{L-band spectra of Be stars. Spectra are normalised and shifted vertically to ease comparison.}
          \label{fig:AtlasL2}
          \end{figure}

          \begin{figure}[b!]
          \centering
          \includegraphics[width=0.95\textwidth]{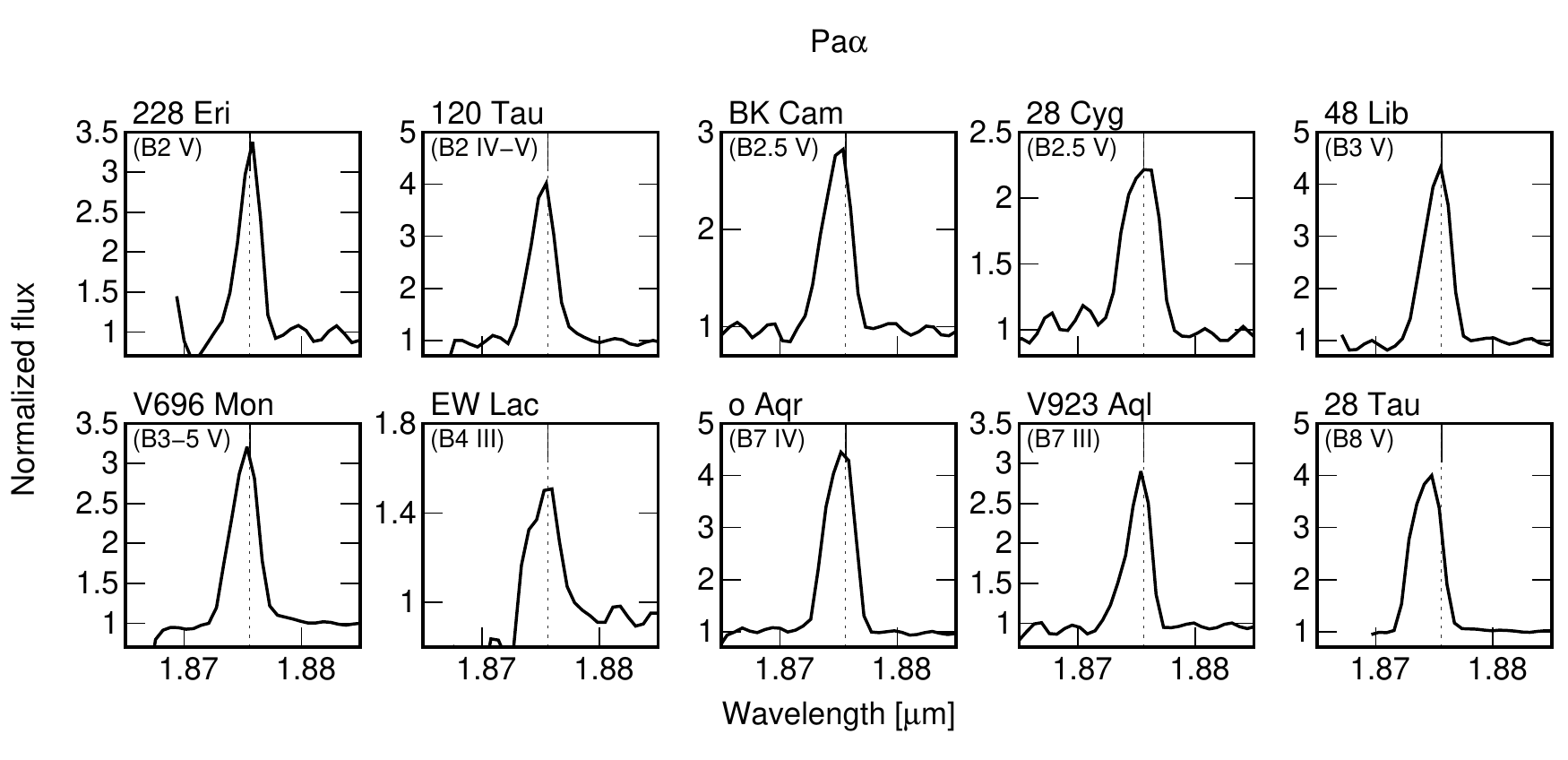}
          \caption{Pa$\alpha$-line profiles.}
          \label{fig:Paalfa}
          \end{figure}
          \clearpage

          \begin{figure} 
          \centering
          \includegraphics[width=0.95\textwidth]{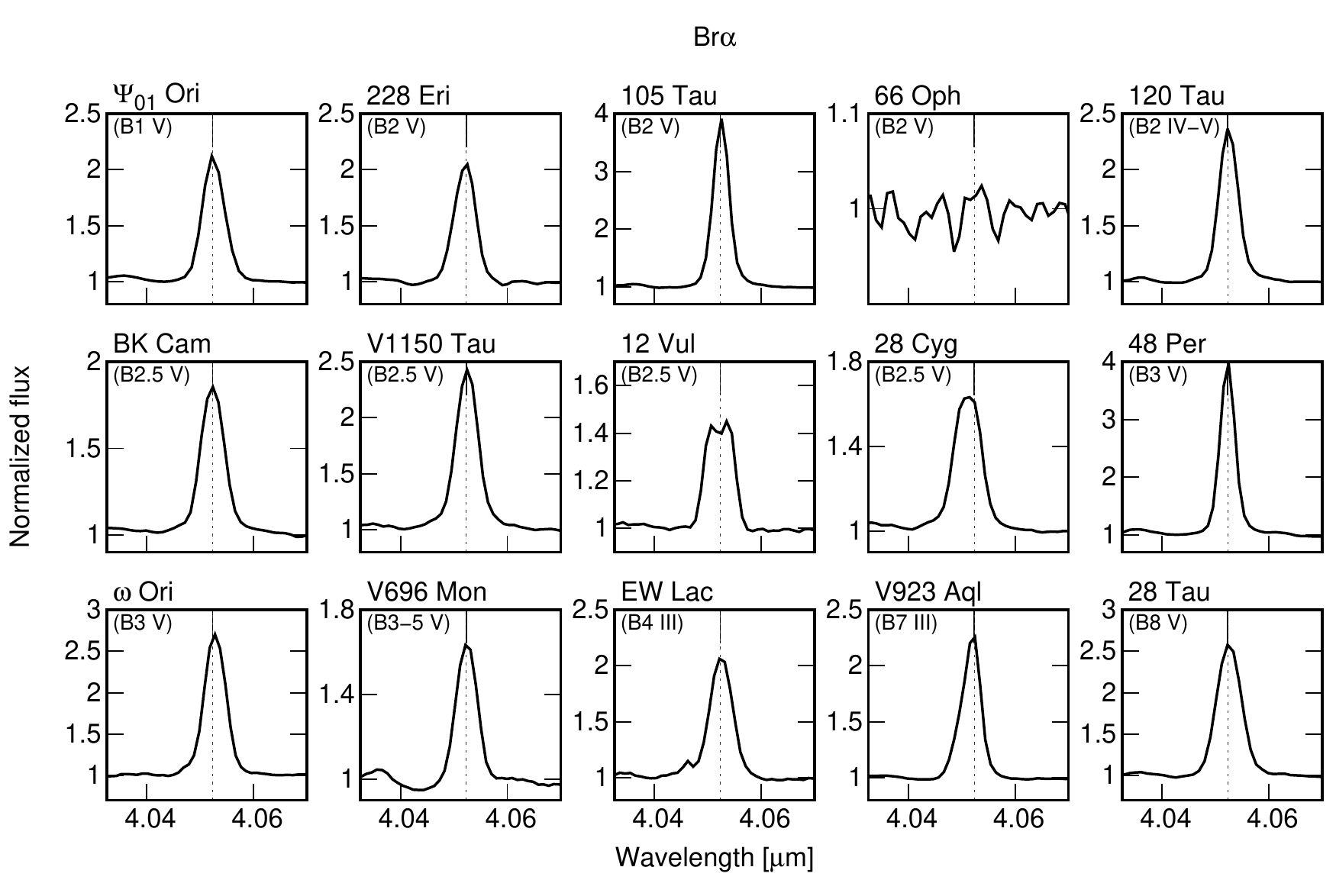}
          \caption{Br$\alpha$-line profiles.}
          \label{fig:Bralfa}
          \end{figure}
          \clearpage

\section{H-line parameters} 

The following tables show the line parameters of each H line for the star sample. The EW are in $\AA$, Fl are in units of 10$^{-13}$\,erg\,cm$^{-2}$\,s$^{-1}\,\AA^{-1}$, and FWHM and $\Delta \mathrm{V}$ are in km\,s$^{-1}$. 

          \begin{table}[h!]
          \caption{$\Psi$\,Ori - EW, Fl, and FWHM of the hydrogen lines.}\label{tabla:PsiOri_Hlines}  
          \centering
\begin{tabular}{ll|ccc} 
          \hline\hline
          Band & Line &  EW     & Fl    & FWHM \\  
          \hline      
          J-Band 
          & Pa$\beta$ &  18.66&  529.86& 310\\
          \hline
          H-Band
          & Br9 & 27.89 & 312.63  & 472 \\
          & Br10  & 14.82 & 173 & 604 \\
          & Br11  & 13.21 & 162.21  & 524 \\
          & Br12  & 11.96 & 153.98  & 507 \\
          & Br13  & 10.41 & 139.14  & 475 \\
          & Br14  & 11.77 & 167.43  & 587 \\
\hline
          \end{tabular}
          \end{table}

          \begin{table}[h!]
          \caption{$\pi$\,Aqr - EW, Fl, FWHM, and $\Delta V$ of the hydrogen lines.}\label{tabla:PiAqr_Hlines}  
          \centering
          \begin{tabular}{ll|cccc} 
          \hline\hline
          Band & Line &  EW     & Fl    & FWHM &$\Delta V$\\  
          \hline      
          J-Band  
          & Pa$\beta$ & 10.12&  41.82&  357 & 212\\
          \hline
          H-Band  
          & Br10  & 6.2 & 9.7 & $\cdots$  & 257 \\
          & Br11  & 5.85  & 9.81  & $\cdots$  & 247 \\
          & Br12  & 5.93  & 10.11 & 455 & 251 \\
          & Br13  & 4.2 & 7.47  & 393 & 272 \\
          & Br14  & 5.54  & 10.37 & 500 & 288 \\
          & Br15  & 4.93  & 9.64  & 473 & 268 \\
          & Br16  & 4.68  & 9.63  & 495 & 293 \\
          & Br17  & 4.88  & 10.39 & 514 & 287 \\
          & Br18  & 4.6 & 10.21 & 491 & 274 \\
          & Br19  & 4 & 9.16  & 490 & 281 \\
          & Br20  & 3.55  & 8.4 & 474 & 274 \\
          & Br21  & 3.96  & 9.52  & 483 & 275 \\
          & Br22  & 2.63  & 6.41  & 480 & 276 \\
          & Br23  & 2.25  & 5.62  & 492 & 269 \\
          \hline
          K-Band   
          & Pf17  & 8.19  & 4.67  & $\cdots$  & 250 \\
          & Pf18  & 4.85  & 2.85  & $\cdots$  & 268 \\
          & Pf19  & 6.73  & 3.99  & $\cdots$  & 307 \\
          & Pf20  & 4.39  & 2.67  & $\cdots$  & 278 \\
          & Pf21  & 6.44  & 3.94  & $\cdots$  & $\cdots$  \\
          & Pf22  & 4.37  & 2.76  & $\cdots$  & $\cdots$  \\
          & Pf23  & 3.69  & 2.37  & $\cdots$  & $\cdots$  \\
          & Pf24  & 3.51  & 2.29  & $\cdots$  & $\cdots$  \\
          & Pf25  & 3.19  & 2.11  & $\cdots$  & $\cdots$  \\
          & Pf26  & 2.16  & 1.45  & $\cdots$  & $\cdots$  \\
          & Pf27-28 & \multicolumn{4}{c}{em}  \\            
          & Br$\gamma$  & 10.51 & 9.93  & 359 & 211\\ 
          \hline
          \end{tabular}
          \end{table}

          \begin{table}[h!]
          \caption{MX\,Pup - EW, Fl, FWHM, and $\Delta V$ of the hydrogen lines.}\label{tabla:MXPup_Hlines}  
          \centering
          \begin{tabular}{ll|cccc} 
          \hline\hline
          Band & Line &  EW     & Fl    & FWHM &$\Delta V$\\  
          \hline      
          J-Band  
          & Pa$\beta$ & 18.52&  39.71 & 191 &98\\
          \hline
          H-Band  
          & Br10  & 15.93&  12.55 & 290 & 116   \\
          & Br11  & 15.79&  13.06 & 310 & 117   \\
          & Br12  & 14.53&  13.13 & 310 & 112   \\
          & Br13  & 13.34&  12.73 & 316 & 144   \\
          & Br14  & 13.72&  13.52 & 339 & 137   \\
          & Br15  & 13.35&  13.49 & 338 & 127   \\
          & Br16  & 12.12&  12.86 & 335 & 106   \\
          & Br17  & 11.28&  12.32 & 345 & 115   \\
          & Br18  & 9.99& 11.19 & 340 & 101   \\
          & Br19  & 8.52& 9.82  & 340 & 113   \\
          & Br20  & 7.21& 8.51  & 336 & 114   \\
          & Br21  & 4.60& 5.71  & 286 & 124   \\
          & Br22  & 4.49& 5.55  & 308 & 111   \\
          & Br23  & 3.25& 4.14  & 295 & 161   \\
          & Br24  & 2.30& 2.99  & 266 & 133   \\
          & Br25  & 2.01& 2.64  & 285 & 140   \\
          & Br26  & 1.63& 2.15  & 286 & 134   \\
          & Br27  & 1.27& 1.71  & 297 & $\cdots$    \\
          & Br28-29 & \multicolumn{4}{c}{em}    \\          
          \hline
          K-Band   
          & Br$\gamma$& 20.90&  8.65  & 229 & 102\\
          \hline
          \end{tabular}
          \end{table}

          \begin{table}[h!]
          \caption{228\,Eri - EW, Fl, and FWHM of the hydrogen lines.}\label{tabla:228Eri_Hlines}  
          \centering
          \begin{tabular}{ll|ccc} 
          \hline\hline
          Band & Line &  EW     & Fl    & FWHM \\  
          \hline      
          J-Band  
          & Pa$\beta$ & 15.97&  343.01& 301\\
          \hline
          H-Band  
          & Br$_9$  & 21.98&  130.5 & 528 \\
          & Br10  & 11.17&  85.2  & 467 \\
          & Br11  & 9.92& 83.61 & 442 \\
          & Br12  & 9.35& 84.11 & 459 \\
          & Br13  & 8.14& 77.73 & 462 \\
          & Br14  & 8.78& 88.57 & 559 \\
\hline
          \end{tabular}
          \end{table}

          \begin{table}[h!]
          \caption{105\,Tau - EW, Fl, and FWHM of the hydrogen lines.}\label{tabla:105Tau_Hlines}  
          \centering
          \begin{tabular}{ll|ccc} 
          \hline\hline
          Band & Line &  EW     & Fl    & FWHM \\  
          \hline      
          J-Band  
          & Pa$\beta$ &27.52&  598.77& 217\\
          \hline
          H-Band  
          & Br$_9$  & 27.72 & 235.55  & 343 \\
          & Br10  & 16.04 & 146.26  & 350 \\
          & Br11  & 14.9  & 143.85  & 343 \\
          & Br12  & 13.93 & 140.94  & 346 \\
          & Br13  & 12.2  & 128.82  & 343 \\
          & Br14  & 12.25 & 139.11  & 384 \\
\hline
          \end{tabular}
          \end{table}

          \begin{table}[h!]
          \caption{V4024\,Sgr - EW, Fl, FWHM, and $\Delta V$ of the hydrogen lines.}\label{tabla:V4024Sgr_Hlines}  
          \centering
          \begin{tabular}{ll|cccc} 
          \hline\hline
          Band & Line &  EW     & Fl    & FWHM &$\Delta V$\\  
          \hline      
          J-Band  
          & Pa$\beta$ & 6.19& 1282.20&  $\cdots$  & 187 \\
          \hline
          H-Band  
          & Br10  & \multicolumn{3}{c}{em}  \\      
          & Br11  & 4.59& 367.65  & $\cdots$  \\
          & Br12  & 4.22& 352.06  & $\cdots$  \\
          & Br13  & 3.40& 320.73  & $\cdots$  \\
          & Br14  & 2.76& 276.09  & $\cdots$  \\
          \hline
          K-Band   
          & Br$\gamma$  & 8.18& 285.61& 295 & 181\\
          \hline
          \end{tabular}
          \end{table}

          \begin{table}[h!]
          \caption{120\,Tau - EW, Fl, and FWHM of the hydrogen lines.}\label{tabla:120Tau_Hlines}  
          \centering
          \hspace{0.07cm}
          \begin{tabular}{ll|ccc} 
          \hline\hline
          Band & Line &  EW     & Fl    & FWHM \\  
          \hline      
          J-Band  
          & Pa$\beta$    &20.46& 1019.65& 308\\
          \hline
          H-Band  
          & Br9 &      25.75& 359.46 &      448    \\
          & Br10     &      16.84& 302.77 &      447    \\
          & Br11     &      14.64& 300.85 &      415    \\
          & Br12     &      14.14& 311.4  &      436    \\
          & Br13     &      13.19& 304.5  &      453    \\
          & Br14     &      13.20& 324.34 &      468    \\
\hline
          \end{tabular}
          \end{table}

          \begin{table}[h!]
          \caption{BK\,Cam - EW, Fl, and FWHM of the hydrogen lines.}\label{tabla:BKCam_Hlines}  
          \centering
          \begin{tabular}{ll|ccc} 
          \hline\hline
          Band & Line &  EW     & Fl    & FWHM \\  
          \hline      
          J-Band  
          & Pa$\beta$  &  15.04&  552.22& 325\\
          \hline
          H-Band  
          & Br9 & 25.4  & 269.77  & 586 \\
          & Br10  & 17.82 & 217.73  & 689 \\
          & Br11  & 12.78 & 172.52  & 473 \\
          & Br12  & 11.38 & 166.64  & 454 \\
          & Br13  & 10.36 & 158.14  & 474 \\
          & Br14  & 10.2  & 165.02  & 514 \\
\hline
          \end{tabular}
          \end{table}

          \begin{table}[h!]
          \caption{V1150\,Tau - EW, Fl, and FWHM of the hydrogen lines.}\label{tabla:V1150Tau_Hlines}  
          \centering
          \begin{tabular}{ll|ccc} 
          \hline\hline
          Band & Line &  EW     & Fl    & FWHM \\ 
          \hline
          H-Band  
          & Br9 & 25.45 & 23.66 & 489 \\
          & Br10  & 17.94 & 19.06 & 497 \\
          & Br11  & 17.45 & 20.53 & 522 \\
          & Br12  & 15.88 & 20.32 & 521 \\
          & Br13  & 13.56 & 18.18 & 504 \\
          & Br14  & 14.08 & 19.66 & 557 \\
          & Br15  & 10.78 & 15.31 & 519 \\
\hline
          \end{tabular}
          \end{table}

          \begin{table}[h!]
          \caption{12\,Vul - EW, Fl, FWHM, and $\Delta V$ of the hydrogen lines.}\label{tabla:12Vul_Hlines}  
          \centering
          \begin{tabular}{ll|cccc} 
          \hline\hline
          Band & Line &  EW     & Fl    & FWHM &  $\Delta V$\\  
          \hline
          H-Band  
          & Br9 & \multicolumn{4}{c}{em}            \\
          & Br10  & 8.52  & 51.11 & 464 & 301 \\
          & Br11  & 7.88  & 53.63 & 417 & 276 \\
          & Br12  & 6.69  & 50.72 & 453 & 266 \\
          & Br13  & 4.96  & 40.44 & 413 & 277 \\
          & Br14  & 4.18  & 36.7  & 395 & 257 \\
          & Br15  & 3.05  & 26.73 & 435 & 264 \\
\hline
          \end{tabular}
          \end{table}

          \begin{table}[h!]
          \caption{28\,Cyg - EW, Fl, FWHM, and $\Delta V$ of the hydrogen lines.}\label{tabla:28Cyg_Hlines}  
          \centering
          \begin{tabular}{ll|cccc} 
          \hline\hline
          Band & Line &  EW     & Fl    & FWHM &  $\Delta V$\\  
          \hline      
          J-Band  
          & Pa$\beta$    & 8.78&  172.70& 386 & $\cdots$ \\
          \hline
          H-Band  
          & Br9 & 18.55 & 95.2  & 571 & $\cdots$  \\
          & Br10  & 7.99  & 51.67 & 576 & 314 \\
          & Br11  & 7.04  & 53.51 & 506 & 345 \\
          & Br12  & 7.08  & 58.72 & 556 & 343 \\
          & Br13  & 5.87  & 52.36 & 540 & 368 \\
          & Br14  & 6.56  & 61.52 & 669 & 348 \\
\hline
          \end{tabular}
          \end{table}

          \begin{table}[h!]
          \caption{48\,Per - EW, Fl, and FWHM of the hydrogen lines.}\label{tabla:48Per_Hlines}  
          \centering
          \begin{tabular}{ll|ccc} 
          \hline\hline
          Band & Line &  EW     & Fl    & FWHM \\  
          \hline
          H-Band
          & Br9 & 19.28&  571.26  & 242 \\
          & Br10  & 11.76&  369.51  & 188 \\
          & Br11  & 10.92&  364.62  & 206 \\
          & Br12  & 9.92& 349.66  & 216 \\
          & Br13  & 9.31& 344.56  & 250 \\
          & Br14  & 8.25& 339.4 & 246 \\
          & Br15  & 6.38& 271.02  & 224 \\
\hline
          \end{tabular}
          \end{table}

          \begin{table}[h!]
          \caption{$\omega$\,Ori - EW, Fl, and FWHM of the hydrogen lines.}\label{tabla:OmOri_Hlines}  
          \centering
          \hspace{0.03cm}
          \begin{tabular}{ll|ccc} 
          \hline\hline
          Band & Line &  EW     & Fl    & FWHM \\  
          \hline      
          J-Band  
          & Pa$\beta$ &  15.75&  554.74& 302\\
          \hline
          H-Band  
          & Br$_9$  & 19.28&  210.59& 389 \\
          & Br10  & 8.96& 112.86& 255 \\
          & Br11  & 8.44& 114.40& 314 \\
          & Br12  & 7.17& 104.80& 209 \\
          & Br13  & 6.21& 95.13&  362 \\
          & Br14  & 4.62& 78.96&  289 \\
          & Br15-20 & \multicolumn{3}{c}{em}      \\
          \hline
          K-Band   
          & Pf18  & 4.18& 15.4  & 279 \\
          & Pf19  & 2.42& 9.26  & 257 \\
          & Pf20  & 2.49& 9.69  & 495 \\
          & Pf21  & 1.14& 4.51  & 342 \\
          & Pf22  & 1.24& 5.01  & 415 \\
          & Pf23  & 0.99& 4.03  & 289 \\
          & Br$\gamma$  & 16.95&  94.73 & 312 \\
          & Br$\delta$  & 20.77&  178.38  & 356 \\
          \hline
          L-Band
          & Hu13  & \multicolumn{3}{c}{em}  \\    
          & Br$\alpha$  & 97.18&  39.33&  304\\ 
          & Hu14  & 15.92&  7.06  & 334\\
          & Hu15  & 14.78&  7.34  & 447\\
          & Hu16  & 12.99&  7.33  & 458\\
          & Hu17  & 7.54& 4.65  & 351\\
          & Pf$\gamma$  & 36.96&  22.78&  342\\ 
          & Hu18  & 4.83& 3.23  & 314\\
          & Hu19  & 4.10& 2.86  & 336\\
          & Hu20  & 3.81& 2.72  & 455\\
          & Hu21  & 2.73& 2.02  & 376\\
          & Hu22  & 2.48& 1.9 & 414\\
          & Hu23  & 1.85& 1.48  & 316\\
          & Hu24  & 1.31& 1.08  & 457\\
          & Hu25  & 1.61& 1.32  & 313\\
          & Hu26-27 & \multicolumn{3}{c}{em}  \\    
          & Pf$\delta$  & 19.90&  20.33&  327\\ 
          \hline
          \end{tabular}
          \end{table}

          \begin{table}[h!]
          \caption{48\,Lib - EW, Fl, FWHM, and $\Delta V$ of the hydrogen lines.}\label{tabla:48Lib_Hlines}  
          \centering
          \begin{tabular}{ll|cccc} 
          \hline\hline
          Band & Line &  EW     & Fl    & FWHM & $\Delta V$\\  
          \hline      
          J-Band  
          & Pa$\beta$ & 16.46 & 599.27& 373 & 233\\
          \hline
          H-Band  
          & Br9 & 18.72&  229.36  & $\cdots$  & 378 \\
          & Br10  & 7.16& 102.19  & $\cdots$  & 400 \\  
          & Br11  & 8.22& 126.03  & $\cdots$  & 356 \\  
          & Br12  & 5.70& 92.23 & $\cdots$  & 434 \\  
          & Br13  & 4.69& 80.35 & $\cdots$  & 423 \\  
          & Br14  & 3.91& 73.99 & $\cdots$  & 488 \\  
          & Br15  & 3.84& 77.32 & $\cdots$  & 478 \\  
          & Br16  & 2.98& 61.05 & $\cdots$  & 519 \\  
          & Br17  & 1.74& 37.03 & $\cdots$  & 498 \\  
          & Br18  & 1.49& 32.35 & $\cdots$  & 519 \\  
          & Br19  & 1.12& 25.01 & $\cdots$  & 508 \\  
          & Br20  & 0.93& 21.18 & $\cdots$  & 540 \\  
          \hline
          K-Band   
          & Pf18  & 11.41&  46.61 & $\cdots$ &  320\\
          & Pf19  & 9.82& 42.6  & $\cdots$ &  262\\
          & Pf20  & 9.18& 40.54 & $\cdots$ &  491\\
          & Pf21  & 4.62& 21.8  & $\cdots$ &  426\\
          & Pf22  & 3.78& 18.4  & $\cdots$ &  407\\
          & Pf23  & 3.52& 17.18 & $\cdots$ &  484\\
          & Pf24  & 2.45& 12.3  & $\cdots$ &  441\\
          & Pf25  & 1.26& 6.44  & $\cdots$ &  $\cdots$ \\
          & Br$\gamma$& 15.39&  104.56  & 377 & 234\\
          & Br$\delta$& 18.63&  194.6 & 520 & 329\\
          & Pa$\alpha$& 108.90& 1201.24 & 374 & $\cdots$ \\
          \hline
          \end{tabular}
          \end{table}

          \begin{table}[h!]
          \caption{V696\,Mon - EW, Fl, and FWHM of the hydrogen lines.}\label{tabla:V696Mon_Hlines}  
          \centering
          \hspace{0.03cm}
          \begin{tabular}{ll|ccc} 
          \hline\hline
          Band & Line &  EW     & Fl    & FWHM \\  
          \hline      
          J-Band  
          & Pa$\beta$ &   16.83&  721.09& 321\\
          \hline
          H-Band  
          & Br9 & 23.79&  312.17  & 531 \\
          & Br10  & 13.51&  214.48  & 542 \\
          & Br11  & 11.49&  198.55  & 389 \\
          & Br12  & 11.18&  206 & 424 \\
          & Br13  & 9.55& 186.56  & 403 \\
          & Br14  & 10.89&  229.88  & 465 \\
\hline
          \end{tabular}
          \end{table}

          \begin{table}[h!]
          \caption{EW\,Lac - EW, Fl, FWHM, and $\Delta V$ of the hydrogen lines.}\label{tabla:EWLac_Hlines}  
          \centering
          \begin{tabular}{ll|cccc} 
          \hline\hline
          Band & Line &  EW     & Fl    & FWHM & $\Delta V$   \\  
          \hline      
          J-Band  
          & Pa$\beta$    & 2.49&   43.31&    $\cdots$  & 570 \\
          \hline
          H-Band  
          & Br10  & 2.56  & 16.08 & $\cdots$ &  $\cdots$  \\
          & Br11  & 2.58  & 17.63 & $\cdots$ &  $\cdots$  \\
          & Br12  & 1.71  & 12.65 & $\cdots$ &  $\cdots$  \\
          & Br13  & 0.28  & 2.15  & $\cdots$ &  $\cdots$  \\
          & Br14-19 & \multicolumn{4}{c}{shell} \\
          \hline
          K-Band   
          & Pf17-22 & \multicolumn{4}{c}{shell}             \\
          & Br$\gamma$  & 6.18  & 16.91 & 796 & 571 \\
          & Br$\delta$  & 14.03 & 56.78 & 793 & 494 \\
          & Pa$\alpha$  & 31.26 & 137.42  & 481 & 292 \\
          \hline
          L-Band
          & Br$\alpha$& 63.13 & 15.52 & 328 & $\cdots$  \\  
          & Hu14  & 16.55 & 4.17  & 805 & 491 \\
          & Hu15  & 14.08 & 4.05  & 700 & 505 \\
          & Hu16  & 9.19  & 2.94  & 785 & 524 \\
          & Hu17  & 9.47  & 3.22  & 471 & 496 \\
          & Pf$\gamma$& 20.65 & 7.16  & 779 & 346 \\  
          & Hu18  & 5.54  & 2.1 & 592 & 474 \\
          & Hu19  & 6.25  & 2.55  & 656 & 479 \\
          & Hu20  & 4.57  & 2 & 652 & 458 \\
          & Hu21  & 3.64  & 1.7 & $\cdots$  & 552 \\
          & Hu22  & 5.06  & 2.45  & $\cdots$  & 628 \\
          & Hu23  & 2.66  & 1.34  & $\cdots$  & 514 \\
          & Hu24  & 1.2 & 0.63  & $\cdots$  & 462 \\
          & Pf$\delta$& 17.14 & 11.65 & 717 & 420 \\  
          \hline
          \end{tabular}
          \end{table}

          \begin{table}[h!]
          \caption{$o$\,Aqr - EW, Fl, FWHM, and $\Delta V$ of the hydrogen lines.}\label{tabla:oAqr_Hlines}  
          \centering
          \begin{tabular}{ll|cccc} 
          \hline\hline
          Band & Line &  EW     & Fl    & FWHM & $\Delta V$   \\   
          \hline      
          J-Band  
          & Pa$\beta$ & 19.77&  603.00& 299 & $\cdots$ \\
          \hline
          H-Band  
          & Br9 & 17.52&  174.2 & 431 & $\cdots$  \\
          & Br10  & 9.44& 102.77  & 437 & 247 \\
          & Br11  & 8.26& 102.29  & 450 & 295 \\
          & Br12  & 7.33& 99.06 & 407 & 303 \\
          & Br13  & 6.84& 98.5  & 514 & 323 \\
          & Br14  & 5.73& 91.13 & 601 & 351 \\
          & Br15  & 5.50& 90.52 & 517 & 352 \\
          & Br16  & 4.04& 67.89 & 713 & 394 \\
          & Br17  & 4.13& 71.76 & 818 & 429 \\
          & Br18  & 3.38& 59.62 & 780 & 429 \\
          & Br19  & 2.28& 40.72 & 602 & 364 \\
          & Br20  & 2.66& 47.72 & 773 & 443 \\
          \hline
          K-Band   
          & Pf17  & \multicolumn{4}{c}{em}  \\          
          & Pf18  & 13.62&  44.43 & 595 & 337 \\
          & Pf19  & 6.23& 22  & 564 & 349 \\
          & Pf20  & 6.38& 22.72 & 644 & 438 \\
          & Pf21  & 5.24& 18.97 & 630 & 365 \\
          & Pf22  & 4.04& 14.92 & 621 & 341 \\
          & Pf23  & 3.52& 13.11 & 708 & 327 \\
          & Pf24  & 2.59& 9.78  & 563 & 458 \\
          & Br$\gamma$  & 19.43&  100.61& 322 & $\cdots$ \\   
          & Br$\delta$  & 21.22&  166.60& 404 & $\cdots$  \\  
          & Pa$\alpha$  & 112.46& 1042.52&  393 & $\cdots$ \\   
          \hline
          \end{tabular}
          \end{table}

          \begin{table}[h!]
          \caption{V923\,Aql - EW, Fl, FWHM, and $\Delta V$ of the hydrogen lines.}\label{tabla:V923Aql_Hlines}  
          \centering
          \begin{tabular}{ll|cccc} 
          \hline\hline
          Band & Line &  EW     & Fl    & FWHM & $\Delta V$   \\   
          \hline      
          J-Band  
          & Pa$\beta$    & 9.21&   17.63& 207 & 362\\
          \hline
          H-Band  
          & Br9 & 8.7 & 5.75  & $\cdots$  & 434 \\
          & Br10-14 & \multicolumn{4}{c}{em}  \\
          \hline
          K-Band   
          & Br$\gamma$  & 9.98  & 3.42  & 275 & $\cdots$  \\
          & Br$\delta$  & 11.93 & 6.07  & 389 &403  \\  
          & Pa$\alpha$  & 60.95 & 32.55 & 333 & $\cdots$  \\
          \hline
          L-Band
          & Br$\alpha$  & 68.12 & 0.97  & 268 & $\cdots$ \\   
          & Hu14  & 13.41 & 0.2 & 372 & 338\\ 
          & Hu15  & 11.95 & 0.22  & 409 & 354\\ 
          & Hu16  & 9.96  & 0.22  & 484 & 370\\ 
          & Hu17  & 8.88  & 0.23  & 401 & 387\\ 
          & Pf$\gamma$  & 21.87 & 0.56  & 256 & 244\\ 
          & Hu18  & 6.85  & 0.2 & 420 & 339 \\
          & Hu19  & 5.3 & 0.17  & 501 & 403 \\
          & Hu20  & 3.83  & 0.13  & $\cdots$  & 398 \\
          & Hu21  & 2.59  & 0.1 & $\cdots$  & 450 \\
          & Hu22  & 4.08  & 0.16  & $\cdots$  & 381 \\
          & Hu23  & 2.25  & 0.1 & $\cdots$  & 373 \\
          & Hu24  & 1.84  & 0.09  & $\cdots$  & $\cdots$  \\
          & Hu25  & 1.84  & 0.09  & $\cdots$  & $\cdots$  \\
          & Hu26  & 0.85  & 0.04  & $\cdots$  & $\cdots$  \\
          & Pf$\delta$  & 21.07 & 1.4 & 335 & 314\\ 
          \hline
          \end{tabular}
          \end{table}

          \begin{table}[h!]
          \caption{28\,Tau - EW, Fl, FWHM, and $\Delta V$ of the hydrogen lines.}\label{tabla:28Tau_Hlines}  
          \centering
          \begin{tabular}{ll|cccc} 
          \hline\hline
          Band & Line &  EW     & Fl    & FWHM & $\Delta V$\\  
          \hline      
          J-Band  
          & Pa$\beta$  &  16.16&  368.68& 383 & $\cdots$ \\
          \hline
          H-Band  
          & Br9 & 14.73 & 102.04  & $\cdots$  & $\cdots$ \\
          & Br10  & 9.43  & 76.86 & $\cdots$  & $\cdots$ \\
          & Br11  & 6.73  & 61.09 & $\cdots$  & $\cdots$ \\
          & Br12  & 6.34  & 61.66 & $\cdots$  & $\cdots$ \\
          & Br13-23 & \multicolumn{3}{c}{shell}         \\
          \hline
          K-Band   
          & Pf17  & 6.13  & 14.15 & $\cdots$  & 366 \\
          & Pf18  & 3.57  & 8.45  & $\cdots$  & 448 \\
          & Pf19  & 3.02  & 7.29  & $\cdots$  & 412 \\
          & Pf20  & 2.4 & 5.97  & $\cdots$  & 544 \\
          & Pf21  & 3.29  & 8.33  & $\cdots$  & 571 \\
          & Br$\gamma$  & 16.26 & 59.77 & 387 & $\cdots$  \\
          & Br$\delta$  & 17.08 & 96.17 & 401 & $\cdots$  \\
          & Pa$\alpha$  & 95.23 & 623.99  & 478 & $\cdots$  \\
          \hline
          L-Band
          & Br$\alpha$  & 99.94 & 5.1 & 348 & $\cdots$ \\ 
          & Hu14  & 17.03 & 0.88  & 619 & 363 \\
          & Hu15  & 13.79 & 0.78  & 601 & 331 \\
          & Hu16  & 12.7  & 0.75  & 685 & 357 \\
          & Hu17  & 9.2 & 0.58  & 535 & 327 \\
          & Pf$\gamma$  & 28.41 & 1.83  & 395 & $\cdots$ \\ 
          & Hu18  & 8.49  & 0.57  & 714 & 440 \\
          & Hu19  & 5.98  & 0.42  & 723 & 450 \\
          & Hu20  & 5.33  & 0.38  & 733 & 420 \\
          & Hu21  & 4.69  & 0.34  & 669 & 393 \\
          & Hu22  & 4.54  & 0.34  & 653 & 428 \\
          & Hu23  & 2.13  & 0.16  & 716 & 414 \\
          & Hu24  & 2.7 & 0.21  & 675 & 363 \\
          & Hu25  & 2.4 & 0.19  & 718 & 401 \\
          & Hu26  & 2.38  & 0.19  & 478 & $\cdots$ \\ 
          & Hu27  & 1.27  & 0.1 & 493 & $\cdots$ \\ 
          & Pf$\delta$  & 26.9  & 2.69  & 691 & 408\\ 
          \hline
          \end{tabular}
          \end{table}

\clearpage

\section{Metallic line parameters} 

        \begin{table}[h!]
        \caption{Metallic line parameters. }\label{tabla:metales1}  
        \centering
        \begin{tabular}{rlcccccccc} 
        \hline\hline
        Element & Line [$\mu$m]  
        & V1075 Sco
        & $\Psi_{01}$\,Ori
        & $\pi$\,Aqr
        & MX Pup
        & 228 Eri
        & 105 Tau
        & 66 Oph
        & V4024 Sgr \\
        \hline 
        O~{\sc i}       &    1.1290    &    na    &     na  &    na     &    na     &    4.41   &    na     &  $\cdots$ &    na     \\
        C~{\sc i}     &    1.1756    &    em    &    1.33   &    2.31   &    0.69   &    1.23   &    2.12   &  $\cdots$ &    0.6    \\
        Mg~{\sc i}      &    1.1828    &  $\cdots$  &    0.53   &  $\cdots$ &  $\cdots$ &    em     &    em     &    em?    &  $\cdots$ \\
        C~{\sc i}       &    1.1899    &  $\cdots$  &  $\cdots$ &  $\cdots$ &  $\cdots$ &    0.27   &    em?    &    -1     &    em?    \\
        He~{\sc i}      &    1.1972    &    -0.42 &    -0.5   &  $\cdots$ &  $\cdots$ &    -0.54  &    -0.47  &    -1.12  &    -0.51  \\
        O~{\sc i}       &    1.2467    &  $\cdots$  &    1.07   &    1.33   &    0.62   &    0.6    &    0.89   &  $\cdots$ &  $\cdots$ \\
        C~{\sc i}       &    1.2565    &  $\cdots$  &  $\cdots$ &    0.16   &  $\cdots$ &    0.38   &    0.67   &  $\cdots$ &  $\cdots$ \\
\hline\hline
        \end{tabular}
        \tablefoot{
        The EW are in $\AA$. em: the line is in emission but difficult to fit; abs: the line is in absorption but difficult to fit; na: spectral range not available. C~{\sc i} 1.6009* is C~{\sc i} 1.6009 $\mu$m + 1.6026 $\mu$m blended. 
        }
        \end{table}

        \clearpage

        \begin{table}
        \caption{Metallic line parameters (continuation of \ref{tabla:metales1}).}\label{tabla:metales2}  
        \centering
        \begin{tabular}{rlccccccc} 
        \hline\hline
        Element & Line [$\mu$m]  
        & 120 Tau
        & BK Cam
        & V1150 Tau
        & 12 Vul
        & 28 Cyg
        & 48 Per
        & $\omega$\,Ori \\
        \hline 
        O~{\sc i}       &    1.1290    & 4.26     &    em     &    na     &    na     &    em     &    na     &  $\cdots$ \\
        C~{\sc i}     &    1.1756    & 1.83     &    em     &    na     &    na     &    0.82   &    na     &  $\cdots$ \\
        Mg~{\sc i}      &    1.1828    & 0.34     &  $\cdots$ &    na     &    na     &    0.49   &    na     &    0.49   \\
        C~{\sc i}       &    1.1899    & em?      &    -1     &    na     &    na     &    0.32   &    na     &  $\cdots$ \\
        He~{\sc i}      &    1.1972    & -0.31      &    -0.84  &    na     &    na     &    -1.07  &    na     &    -0.75  \\
        O~{\sc i}       &    1.2467    & 1.06     &    0.98   &    na     &    na     &    0.4    &    na     &    em   \\
        C~{\sc i}       &    1.2565    & $\cdots$ &  $\cdots$ &    na     &    na     &  $\cdots$ &    na     &  $\cdots$ \\
\hline\hline
        \end{tabular}
        \end{table}

        \clearpage

        \begin{table}
        \caption{Metallic line parameters (continuation of \ref{tabla:metales1}).}\label{tabla:metales3}  
        \centering
        \begin{tabular}{rlccccccc} 
        \hline\hline
        Element & Line [$\mu$m]  
        &48 Lib
        &V696 Mon
        &EW Lac
        &$o$ Aqr
        &V923 Aql
        &28 Tau
        &HD 171623\\
        \hline 
        O~{\sc i}      &    1.1290    & 6.08    &    5.03   &  $\cdots$ &   em    &    em     &    em     &  $\cdots$  \\
        C~{\sc i}      &    1.1756    & 1.11    &    3.83   &  $\cdots$ &   0.71  &  $\cdots$ &    0.79   &  $\cdots$  \\
        Mg~{\sc i}     &    1.1828    & em      &    1.94   &  $\cdots$ &   0.5   &  $\cdots$ &    em     &  $\cdots$  \\
        C~{\sc i}      &    1.1899    & $\cdots$    &    1.36   &  $\cdots$ &   0.29  &  $\cdots$ &  $\cdots$ &  $\cdots$  \\
        He~{\sc i}     &    1.1972    & -0.96       &    -0.82  &    -0.91  & $\cdots$  &  $\cdots$ &  $\cdots$ &  $\cdots$  \\
        O~{\sc i}      &    1.2467    & 0.75    &    1.38   &    0.14   &   0.31  &    0.33   &    0.42   &  $\cdots$  \\
        C~{\sc i}      &    1.2565    & $\cdots$    &    0.86   &  $\cdots$ &   0.31  &  $\cdots$ &  $\cdots$ &  $\cdots$  \\
\hline\hline
        \end{tabular}
        \end{table}

\end{appendix}

\end{document}